\documentclass{aastex631}

\usepackage{bm}
\usepackage{multirow}
\turnoffeditone

\accepted{August 23, 2022}

\shorttitle{AASTeX v6.3.1 Sample article}
\shortauthors{Kinoshita et al.}
\graphicspath{{./}{figures/}}

\begin{document}

\title{MHD simulations of dense core collision}

\author{Shinichi. W. Kinoshita}
\affiliation{Department of Astronomy, the University of Tokyo \\
7-3-1 Hongo Bunkyo,113-0033 \\
 Tokyo, Japan}
\affiliation{National Astronomical Observatory of Japan \\
NINS, 2-21-1 Osawa, Mitaka\\
Tokyo 181-8588, Japan}.

\author{Fumitaka Nakamura}
\affiliation{Department of Astronomy, the University of Tokyo \\
7-3-1 Hongo Bunkyo,113-0033 \\
 Tokyo, Japan}
\affiliation{National Astronomical Observatory of Japan \\
NINS, 2-21-1 Osawa, Mitaka\\
Tokyo 181-8588, Japan}.
\affiliation{The Graduate University for Advanced Studies (SOKENDAI)\\
2-21-1 Osawa, Mitaka\\
Tokyo 181-0015, Japan}.



\begin{abstract}
We investigated the effect of magnetic fields on the collision process between dense molecular cores. We performed three-dimensional magnetohydrodynamic simulations of collisions between two self-gravitating cores using the {\tt Enzo} adaptive mesh refinement code. The core was modeled as a stable isothermal Bonnor–-Ebert (BE) sphere immersed in uniform magnetic fields. Collisions were characterized by the offset parameter $b$, Mach number of the initial core $\mathcal{M}$, magnetic field strength $B_{0}$, and angle $\theta$ between the initial magnetic field and collision axis. For head-on ($b = 0$) collisions, one protostar was formed in the compressed layer. The higher the magnetic field strength, the lower the accretion rate. For models with $b = 0$ and $\theta = 90^{\circ}$, the accretion rate was more dependent on the initial magnetic field strength compared with $b = 0$ and $\theta = 0^{\circ}$ models. For off-center ($b = 1$) collisions, the higher specific angular momentum increased; therefore, the gas motion was complicated. In models with $b = 1$ and $\mathcal{M} = 1$, the number of protostars and gas motion highly depended on $B_{0}$ and $\theta$. For models with $b = 1$ and $\mathcal{M} = 3$, no significant shock-compressed layer was formed and star formation was not triggered.

\end{abstract}



\section{Introduction}
There is increasing observational evidence that cloud–cloud collisions trigger star and star cluster formation in the Milky Way (e.g., \citealp{1976ApJ...209..466L}; \citealp{1986ApJ...310L..77S}; \citealp{1994ApJ...429L..77H}; \citealp{2011ApJ...738...46T}; \citealp{2021PASJ...73S.300K}). Therefore, collisions between clouds are an important process in structure formation in the ISM.
Theoretical calculations have also demonstrated that cloud-cloud collisions can provide a viable mechanism for triggering star formation (e.g., \citealp{2000ApJ...536..173T}).
Converging flows are likely to produce dense gas clumps that tend to be gravitationally unstable, and are potential precursors to massive stars and star clusters. Observational evidence of triggered star formation by cloud-cloud collisions comes mainly from giant molecular cloud-scale events that produce massive stars and clusters (e.g., \citealp{2016ApJ...820...26F}). Therefore, many numerical studies to date have focused on collisions between high-mass clouds (e.g., \citealp{1992Natur.359..207C}; \edit1{\citealp{1992PASJ...44..203H}}; \edit1{\citealp{2000ApJ...536..173T}}; \citealp{2013ApJ...774L..31I}; \edit1{\citealp{2014ApJ...792...63T}}; \citealp{2015ApJ...811...56W}; \edit1{\citealp{2018PASJ...70S..53I}}; and \edit1{\citealp{2018PASJ...70S..58T}}; and \edit1{\citealp{2018PASJ...70S..57W}}; \edit1{\citealp{2021PASJ...73S.405F}}).

Collisions between smaller gas structures can also occur in turbulent molecular clouds. A velocity dispersion of approximately a few $\rm km ~s^{-1}$ is often found in star forming regions and in the interiors of giant molecular clouds. Through turbulent motion, clouds can become filamentary and flocculent, and their gas can be locally compressed to form a dense molecular cloud core. In such turbulent clouds, collisions between dense cores are possible because of the velocity dispersion between them. \edit1{Some theoretical studies (e.g., \citealp{Inutsuka_1997}) have suggested that coalescence among dense cores can occur after the fragmentation of their parental filamentary cloud, depending on the dynamic state in the filament.}

We can roughly estimate the geometrical collision timescale for dense cores. Cores’ mean free path for a geometrical encounter is given by 
\begin{equation}
 \label{eq:lambda}
 \lambda = \frac{1}{\sigma n_{\rm c}}. 
\end{equation}
Here, we assume that cores with a collision cross-section of $\sigma$ are distributed uniformly in the cloud with a core number density of $n_{\rm c}$. 
Then, the collision time scale can be estimated as: 
\begin{equation}
 \label{eq:lambda/v}
 \tau_{\rm coll} = \frac{\lambda}{\langle v \rangle} = \frac{1}{2\pi r_{\rm c}^{2} n_{\rm c} \langle v \rangle}. 
\end{equation}
In the above estimation, we assumed the collision of identical cores and replaced $\sigma$ with $2\pi r_{\rm c}^{2}$, where $r_{\rm c}$ is the radius of the core that experiences collisions; $\langle v \rangle$ is the mean velocity of the core with respect to that of the parent cloud; and $\tau_{\rm coll}$ is expressed as follows: 
\begin{equation}
 \label{eq:collision_timescale}
 \tau_{\rm coll}\sim 0.3 \left(\frac{r_{\rm c}}{0.05~\rm pc}\right)^{-2}\left(\frac{n_{\rm c}}{\rm 200~pc^{-3}}\right)^{-1}
 \left(\frac{\langle v \rangle}{\rm km~s^{-1}}\right)^{-1}~\rm Myr,
\end{equation}

As an observational example, we use the core sample of the Orion Nebula Cluster (ONC) region to estimate the collision timescale, $\tau_{\rm coll}$, in GMCs.
This area was recently observed by CRAMA-NRO \citep{2018ApJS..236...25K}. Using CARMA-NRO C$^{18}$O $(J = 1–0)$ data, \citet{2021ApJ...910L...6T} identified approximately 200 dense cores in the filamentary region of ONC ($\sim1\rm ~pc^{-3}$ area), which gives a number density of $n_{\rm c}\sim 200 ~\rm pc^{-3}$. If we adopt $r_{\rm c} = 0.05~\rm pc$ and $\langle v \rangle = 1~\rm km~s^{-1}$ as representative values, we can derive the collision time $\tau_{\rm coll}\sim 0.3 ~\rm Myr $ using Equation \ref{eq:collision_timescale}. \citet{2021ApJ...910L...6T} derived the core lifetime in Orion A of $\sim5~t_{\rm ff}$ for starless cores with densities of $\rho_{\rm 0}\sim 10^{4}-10^{5}~ \rm cm^{-3}$, where $t_{\rm ff} = (3 \pi /[32 G \rho_{\rm 0}])^{1 / 2}$ is free-fall time of the core. Thus, the core lifetime is evaluated as $\sim5~t_{\rm ff}\simeq 0.5-1.5~\rm Myr$
Therefore, the average core experienced several collisions before a star is created. In the denser region, more frequent collisions would occur. Thus, in real clouds, core collisions are expected to influence the evolution of the cores. \edit1{\citet{2013ApJ...769...23H} also suggested that core coagulation is one of the dominant physical processes to determine the form of dense cores’ mass function (CMF).}


Collisions between dense clumps or cores have been studied numerically for decades (e.g., \citealp{1992Natur.359..207C}, \citealp{1995MNRAS.277..727W}; \citealp{2000A&AS..142..165M} \citealp{2001A&A...379.1123M}). These studies have investigated the evolution of several orders of magnitude in density and protostellar fragmentation on small scales.
In recent years, simulations with higher spatial resolutions have been developed, and more precise numerical calculations have been performed. For instance, \citet{2007MNRAS.378..507K} investigated the case of colliding gravitationally stable clumps ($\geq 10M_{\odot}$) using smoothed particle hydrodynamic (SPH) simulations. They showed that collisions produce shock-compressed layers that fragment into filaments depending on parameters such as offset parameter $b$ and Mach number $\mathcal{M}$. In their simulations, protostellar objects condensed from these filaments and accreted. \citet{2015AN....336..695A} presented 3D hydrodynamic simulations of two rotating-core ($8M_{\odot}$) collisions. In their simulation, high-density objects that may have evolved into protostars were formed. Previous studies, primarily using hydrodynamic simulations, have demonstrated that core collisions have a significant impact on the star formation process.

\edit1{On the other hand, some observations suggest magnetic field significantly influences core dynamic evolution (e.g., \citealp{2004mim..proc..123C}). However, previous numerical studies of dense core collisions have not considered magnetic filelds.} In this paper, we consider the effect of magnetic fields on the \edit1{dense core} collision process and focus on their observational features. Although previous studies did not include magnetic fields, observations suggest that the magnetic field significantly influences core dynamic evolution (e.g., \citealp{2004mim..proc..123C}). \edit1{Here, we address collisions between $\sim 4M_{\odot}$ dense cores immersed in magnetic fields.}

The remainder of this paper is organized as follows. In Section \ref{sec:method}, we describe the numerical, model, and analysis methods used. In Section \ref{sec:result}, we present important features of our simulation results. In Section \ref{sec:discussion}, the results are discussed. Section \ref{sec:conclusion} presents our main conclusions.

\section{Methods}

\begin{figure*}
  
    \begin{tabular}{c}

      \begin{minipage}{0.5\hsize}
        \begin{center}
          \includegraphics[clip, width=8.0cm]{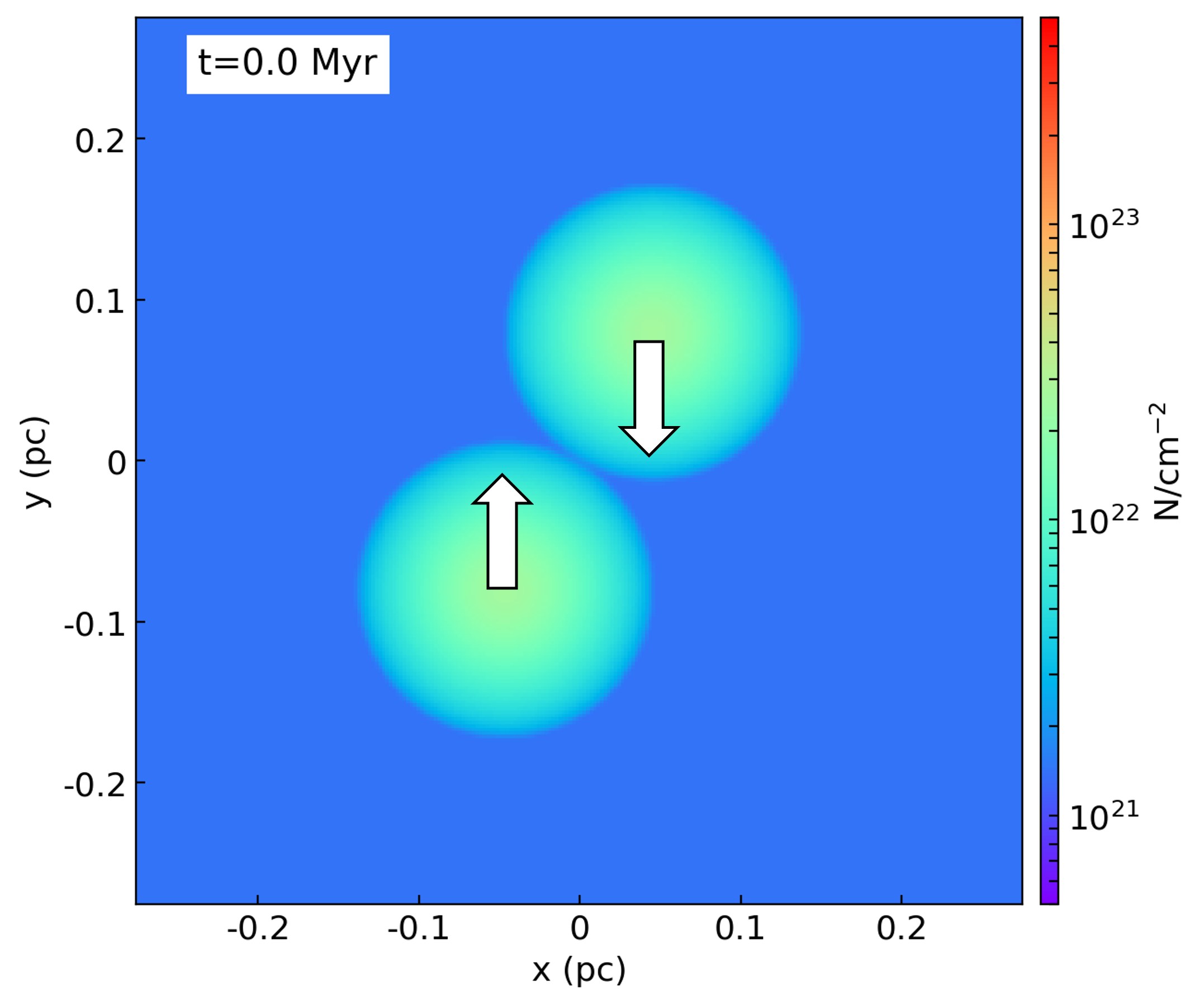}
        \end{center}
      \end{minipage}

      \begin{minipage}{0.5\hsize}
        \begin{center}
          \includegraphics[clip, width=8.0cm]{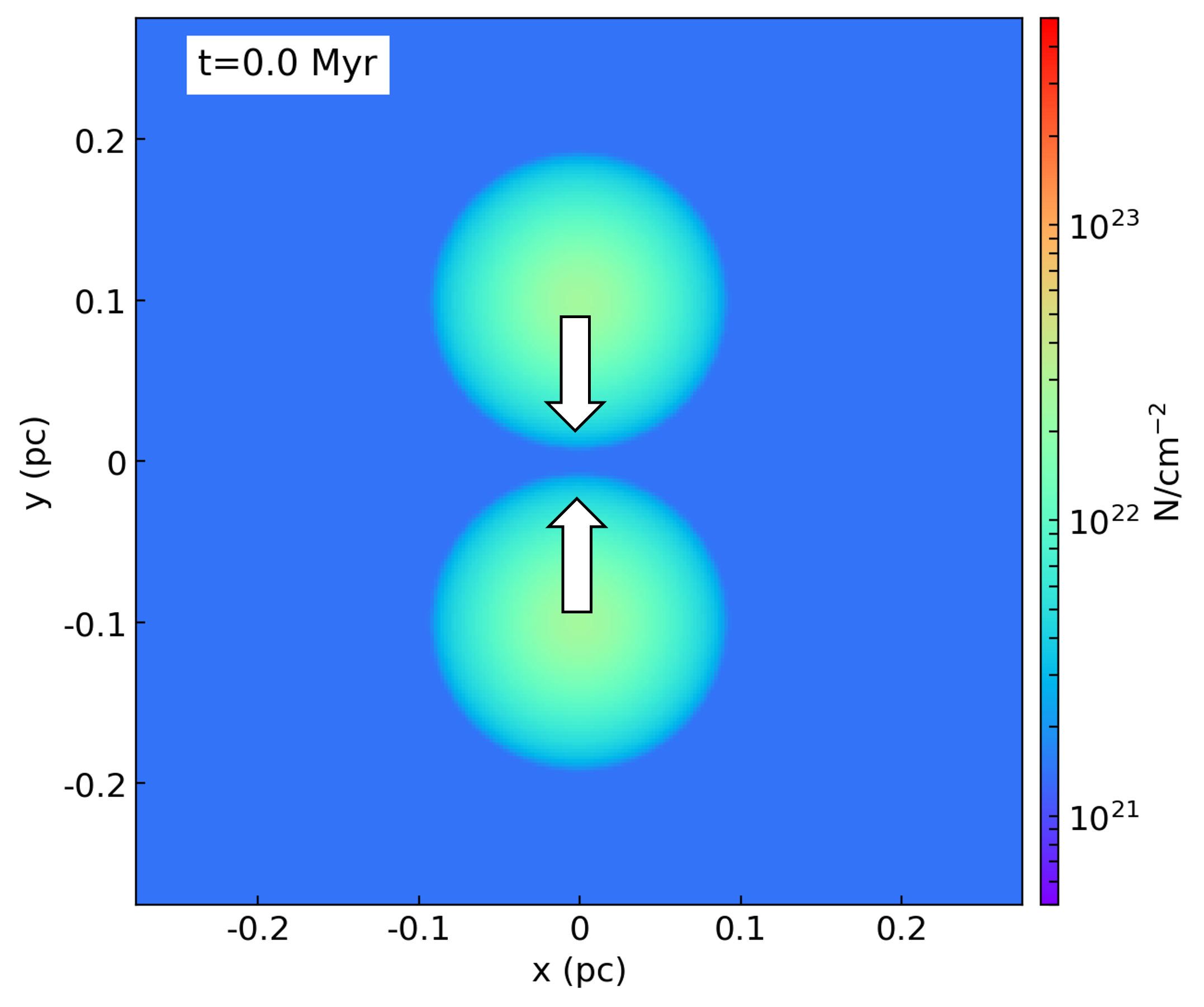}
        \end{center}
      \end{minipage}

       \end{tabular}

    \vskip5pt  
    \caption{Initial conditions with the mass surface density shown. White arrows indicate the direction of each core's motion. \textit{Left panel}: Head-on collision cases (b = 0).
    \textit{Right panel}: off-center collision cases (b = $r_{\rm c}$)
\label{fig:initial_condition}}

\end{figure*}
\label{sec:method}
\subsection{Initial conditions}
In the simulations, we considered the collision between two cores of equal mass. For the initial core, we set a stable Bonnor--Ebert (BE) sphere (\citealp{1955ZA.....37..217E}, \citealp{1956MNRAS.116..351B}; \citealp{2001Natur.409..159A}). BE spheres are hydrostatic solutions of self-gravitating isothermal clouds confined by an external pressure. Some observations have indicated that the density profiles of dense cores can be approximated by those of BE spheres (e.g., \citealp{2001ApJ...557..193E}). The initial BE sphere in our simulation has a radius of $r_{\rm c} = $\edit1{0.1}$~\rm pc$, central density of $\rho_{\rm c} = 10^{5} ~\rm cm^{-3}$, and a temperature of $T_{\rm c} = $\edit1{20}$ ~\rm K$, giving a dimensionless radius of $\xi\equiv(4\pi G\rho_{\rm c}/c_{\rm s}^{2})^{1/2}r_{\rm c} = $\edit1{6.0} and mass of $M_{\rm c} = $\edit1{3.7}$~M_{\odot}$, where $c_{\rm s}\simeq$ \edit1{0.27}$~\rm km~s^{-1}$ is the sound speed in the core. This dimensionless radius is \edit1{below} the critical value of $\xi_{\rm crit} = 6.45$. The free-fall time $t_{\rm ff}$ corresponding to the core central density $\rho_{\rm c}$ was $t_{\rm ff} = (3 \pi /[32 G \rho_{\rm c}])^{1 / 2}\simeq$ 0.1 Myr. 

The ambient medium was uniform and initialized to satisfy the pressure balance at the core boundary. The density contrast between the cloud surface and ambient ISM gas was $\chi_{\rm c} = 10$. We assumed a mean molecular weight of $\mu = 2.3$ and the adiabatic index was set to $\gamma = 1.00001$ for an approximate isothermal assumption. This initial setup is similar to that of \citet{2021ApJ...921..150K}, who examined star formation triggered by large-scale shocks.

We considered both head-on and off-center collisions. Figure \ref{fig:initial_condition} shows the initial setup. Our simulation volume was $\sim$\edit1{0.6} pc sided cubic. \edit1{Two cores had initial separations of their centers of $\sim 2~r_{\rm c}$ in the $y$-direction. For head-on collisions, in the $x$ direction, they had no separations. We defined the offset parameter $b$ as the ratio of the impact parameter of the collision to the core radius. For off-center collisions, the colliding cores were displaced by $b = 1$ along the $x$-direction. For both collisions, the two cores had identical pre-collision velocities $v_{\rm c}$ along the $y$-direction. We explored $v_{\rm c} = 1.0 ~c_{\rm s} $ and $v_{\rm c} = 3.0~c_{\rm s} $ cases corresponding to $\mathcal{M} = 1,3$, where a Mach number of $\mathcal{M}$ is the ratio of the collision velocity to the sound speed in the initial core. Thus, relative collision velocities are $v_{\rm rel} = 2.0~c_{\rm s} $ or $v_{\rm rel} = 6.0~c_{\rm s} $ ($\mathcal{M}=2,6$).}

\edit1{The simulation box is initialized with a uniform magnetic field directed at an angle $\theta$ relative to the collision axis of the cores. We considered the initial magnetic fields parallel to the $y$-direction ($\theta = 0^{\circ}$) and parallel to the $z$-direction ($\theta = 90^{\circ}$) to explore the effects of magnetic field orientations.} 
\edit1{For cases with $b = 1$ and $\mathcal{M} = 1$, we additionally considered models with the oblique magnetic field in the y–z plane with an angle of $\theta = 45^{\circ}$ with respect to the $z$-axis.}

We selected three magnetic field strengths: $B_{0} = $\edit1{10}$~\mu \rm G$, $B_{0} = $\edit1{30}$~\mu \rm G$, and $B_{0} = $\edit1{50}$~\mu \rm G$. When we define the mass-to-flux ratio normalized to the critical value, $\mu\equiv(M/\Phi)/(M/\Phi)_{\rm crit}$, the magnetic field strengths correspond to $\mu=$\edit1{6.0, 2.0, and 1.2} for the $B_{0} = $\edit1{10, 30, and 50}$~\mu \rm G$ models, respectively. Here, we adopted $(M/\Phi)_{\rm crit}=1/2\pi\sqrt{G}$ \citep{1978PASJ...30..671N}. Thus, for $B_{0} = $\edit1{10}$~ \mu \rm G$ cases, the magnetic field has little effect, whereas for $B_{0} = $\edit1{50}$~\mu \rm G$ cases, the magnetic energy is comparable to the gravitational energy. 

Properties of initial conditions are listed in Table \ref{tab:Initial Simulation parameter}. A summary of the models is presented in Table \ref{tab:Explored Parameter Space}. In general, the simulations were performed for $\sim 10~t_{\rm ff}\simeq$ 1 Myr. If the core material touched the boundary of the simulation box, we terminated the simulation. 


\begin{deluxetable*}{ccccc}
\tablecaption{Initial simulation parameters}
\tablewidth{0pt}
\tablehead{\hspace{4.0cm}
& \hspace{4.0cm}       &      Core \hspace{4.0cm} & Ambient \hspace{4.0cm}\\
}
\startdata
$r_{\rm c}$ $^{\rm a}$        & (pc)          &  0.1  & ... \\
$\xi$ $^{\rm b}$       &               &  6.0  & ... \\
$c_{\rm s}$ $^{\rm c}$ & ($\rm km~s^{-1}$) & 0.27   &  0.85\\
$M_{\rm c}$ $^{\rm d}$ & ($M_{\odot}$) &  3.7  & ... \\
$\rho_{\rm c}$ $^{\rm e}$ & ($\rm cm^{-3}$) & $10^{5}$ & ... \\
$t_{\rm ff}$ $^{\rm f}$           &    (Myr)           & 0.1   & ... \\
$B_{0}$  $^{\rm g}$ & ($\mu G$) &\multicolumn{2}{c}{ (10, 30, 50) $^{\rm h}$}
\enddata
\tablenotetext{\tiny \rm{a}}{Core radius.}
\tablenotetext{\tiny \rm{b}}{Dimensionless radius of the core.}
\tablenotetext{\tiny \rm{c}}{Sound speed of gas.}
\tablenotetext{\tiny \rm{d}}{Core mass.}
\tablenotetext{\tiny \rm{e}}{Central density of the core.}
\tablenotetext{\tiny \rm{f}}{Free-fall time ($3 \pi /[32 G \rho_{\rm c}])^{1 / 2}$.}
\tablenotetext{\tiny \rm{g}}{Magnetic fields.}
\tablenotetext{\tiny \rm{h}}{Applied to the entire computation box.}
\label{tab:Initial Simulation parameter}
\end{deluxetable*}

\begin{deluxetable*}{lccccccc}
\tablecaption{Explored parameter Space}
\tablewidth{0pt}
\tablehead{
Model name $^{\rm a}$ & $b$ $^{\rm b}$ & $\mathcal{M}$ $^{\rm c}$ &  $v_{\rm c}$ $^{\rm d}$   & $\theta$ $^{\rm e}$  &  \colhead{$B_{0}$ $^{\rm f}$} &$\mu$ $^{\rm g}$ & $\beta$ $^{\rm h}$ \\
& ($r_{\rm c}$) &  & ($\rm km ~s^{-1}$)  & (degrees) & \colhead{($\mu G$)} & &
}
\startdata
{\tt b0-M1-By10}&  0  & 1.0          &  0.27    &$0$ & 10 & 6.0  & 7.0        \\  
{\tt b0-M1-By30}&  0  & 1.0          &  0.27    &$0$ & 30 & 2.0  & 0.8        \\
{\tt b0-M1-By50}&  0  & 1.0          &  0.27    &$0$ & 50 & 1.2  & 0.3        \\
\hline
{\tt b0-M1-Bz10}&  0  & 1.0          &  0.27    &$90$& 10 & 6.0  & 7.0        \\  
{\tt b0-M1-Bz30}&  0  & 1.0          &  0.27    &$90$& 30 & 2.0  & 0.8        \\
{\tt b0-M1-Bz50}&  0  & 1.0          &  0.27    &$90$& 50 & 1.2  & 0.3        \\
\hline
{\tt b0-M3-By10}&  0  & 3.0          &  0.80    &$0$ & 10 & 6.0  & 7.0        \\  
{\tt b0-M3-By30}&  0  & 3.0          &  0.80    &$0$ & 30 & 2.0  & 0.8        \\
{\tt b0-M3-By50}&  0  & 3.0          &  0.80    &$0$ & 50 & 1.2  & 0.3        \\
\hline
{\tt b0-M3-Bz10}&  0  & 3.0          &  0.80    &$90$& 10 & 6.0  & 7.0        \\  
{\tt b0-M3-Bz30}&  0  & 3.0          &  0.80    &$90$& 30 & 2.0  & 0.8        \\
{\tt b0-M3-Bz50}&  0  & 3.0          &  0.80    &$90$& 50 & 1.2  & 0.3        \\
\hline
\hline
{\tt b1-M1-By10}&  1  & 1.0          &  0.27    &$0$ & 10 & 6.0  & 7.0        \\  
{\tt b1-M1-By30}&  1  & 1.0          &  0.27    &$0$ & 30 & 2.0  & 0.8        \\
{\tt b1-M1-By50}&  1  & 1.0          &  0.27    &$0$ & 50 & 1.2  & 0.3        \\
\hline
{\tt b1-M1-Bz10}&  1  & 1.0          &  0.27    &$90$& 10 & 6.0  & 7.0        \\  
{\tt b1-M1-Bz30}&  1  & 1.0          &  0.27    &$90$& 30 & 2.0  & 0.8        \\
{\tt b1-M1-Bz50}&  1  & 1.0          &  0.27    &$90$& 50 & 1.2  & 0.3        \\
\hline
{\tt b1-M1-Bob10}&  1  & 1.0          &  0.27    &$45$ & 10 & 6.0  & 7.0        \\  
{\tt b1-M1-Bob30}&  1  & 1.0          &  0.27    &$45$ & 30 & 2.0  & 0.8        \\
{\tt b1-M1-Bob50}&  1  & 1.0          &  0.27    &$45$ & 50 & 1.2  & 0.3        \\
\hline
{\tt b1-M3-By10}&  1  & 3.0          &  0.80    &$0$ & 10 & 6.0  & 7.0        \\  
{\tt b1-M3-By30}&  1  & 3.0          &  0.80    &$0$ & 30 & 2.0  & 0.8        \\
{\tt b1-M3-By50}&  1  & 3.0          &  0.80    &$0$ & 50 & 1.2  & 0.3        \\
\hline
{\tt b1-M3-Bz10}&  1  & 3.0          &  0.80    &$90$& 10 & 6.0  & 7.0        \\  
{\tt b1-M3-Bz30}&  1  & 3.0          &  0.80    &$90$& 30 & 2.0  & 0.8        \\
{\tt b1-M3-Bz50}&  1  & 3.0          &  0.80    &$90$& 50 & 1.2  & 0.3        \\
\enddata
\tablenotetext{\tiny \rm{a}}{Model names.
"{\tt b0}" and "{\tt b1}" refer to the head-on collision ($b = 0$) and off-center collision ($b = 1$). "{\tt M1}" and "{\tt M3}" refer to the pre-collision velocity $\mathcal{M} = 1$ and $\mathcal{M} = 3$. "{\tt Bz10}", "{\tt Bz30}", and "{\tt Bz50}" refer to the initial magnetic field strength $B_{\rm z} = 10, 30$, and $50~\mu G$. "{\tt By10}", "{\tt By30}", and "{\tt By50}" refer to $B_{\rm y} = 10, 30$, and $50~\mu G$. "{\tt Bob10}", "{\tt Bob30}", and "{\tt Bob50}" refer to the initial magnetic field strength $B_{0} = 10, 30$, and $50~\mu G$ with $\theta = 45^{\circ}$. }
\tablenotetext{\tiny \rm{b}}{Impact parameter.}
\tablenotetext{\tiny \rm{c}}{Mach number at which initial core moves.}
\tablenotetext{\tiny \rm{d}}{Initial speed of the core.}
\tablenotetext{\tiny \rm{e}}{Angle between the initial magnetic field and the collision axis}
\tablenotetext{\tiny \rm{f}}{Initial magnetic field strength.}
\tablenotetext{\tiny \rm{g}}{Mass-to-flux ratio normalized to the critical value: $\mu\equiv(M/\Phi)/(M/\Phi)_{\rm crit}$.}
\tablenotetext{\tiny \rm{h}}{Thermal-to-magnetic pressure ratio of the initial core surface: $\beta = 8\pi\rho_{0}c_{0}^{2}/B^{2}$.}
\label{tab:Explored Parameter Space}
\end{deluxetable*}

\subsection{Numerical code}
In this study, simulations were conducted using {\tt Enzo}\footnote{http://enzo-project.org (v.2.6)}, a magnetohydrodynamic (MHD) adaptive mesh refinement (AMR) code \citep{2014ApJS..211...19B}. The ideal MHD equations were solved using a Runge--Kutta second-order-based MUSCL solver, including Dedner MHD (\citealp{DEDNER2002645}; \citealp{2008ApJS..176..467W}). The Riemann problem was solved using the Harten-–Lax--van Leer (HLL) method, while the reconstruction method for the MUSCL solver was a piecewise linear model (PLM). Readers are referred to \citet{2014ApJS..211...19B} for more details.

The simulation had a top-level root grid of $256^{3}$ with \edit1{five} additional levels of refinement, corresponding to a maximum resolution of \edit1{8,192}$^{3}$. We used the Jeans criterion to prevent spurious fragmentation \citep{1997ApJ...489L.179T}. We
adopted the limit in which the Jeans length does not fall below
eight cells: $\Delta x < \lambda_{j}/8$, where $\lambda_{j} = \pi^{1/2}c/(G\rho)^{1/2}$ is the Jeans length. Refinement was allowed until the finest resolution reached $\Delta x_{\min} = L/$\edit1{8,192}$\simeq$ \edit1{6.7}$\times 10^{-5} ~\rm pc$, where the local density reached $n_{\rm crit}\simeq$ \edit1{3.2}$\times 10^{9} ~\rm cm^{-3}$. When the Jeans length became unresolved, instead of creating another AMR level, we used the sink particle technique.
When sink particles are formed, their density is assigned to the finest grids using a second-order cloud-in-cell interpolation technique \citep{1988csup.book.....H}, and they move through the grid via gravitational interactions with the surrounding gas and other particles \citep{2014ApJS..211...19B}. 

\subsection{Analysis}

We used the following analysis methods to study the core-core collision process: 

\subsubsection{Color variable}
\label{sec:color}
As in previous studies (e.g., \citealp{1995ApJ...454..172X}; \citealp{2021ApJ...921..150K}) to quantitatively follow the evolution of the core material, we solved the following additional advection equation:
\begin{equation}
 \label{eq:color_variable}
 \frac{(\partial \rho C_{i})}{\partial t}+\nabla \cdot(\rho C_{i} \bm{v}) = 0,
\end{equation}

where $C_{i}$ represents the set of two Lagrangian tracers. The subscript $i$ denotes the two initial core materials: $i = 1$ and 2. Initially, for one core, we defined $C_{1} = 1$ and $C_{2} = 0$, whereas for the other core, we defined $C_{1} = 0$ and $C_{2} = 1$. The ambient gas was labeled $C_{1},C_{2} = 0$. During core-core collision, the core material is mixed with the ambient gas, resulting in regions with $0 < C_{i} < 1$. In any zone, the density of the core material was $\rho_{i} = \rho C_{i}$. This configuration allows us to investigate the evolution of the core material.

\subsubsection{Derivation of polarized emissions}
To study the influence of collisions on magnetic field orientations, we derived the polarized emissions $\bm{p}$ using a method based on previous works (e.g., \citealp{2016ApJ...829...84C}; \citealp{2017ApJ...835..137W}).
Using a Cartesian coordinate system, where the $z$-axis corresponds to the north and the $x$-axis is parallel to the line of sight, the pseudo-vector $\bm{p}$ is defined as

\begin{equation}
 \label{eq:pseudo-vector}
 \bm{p} = (p \sin \chi) \hat{\bm{x}}+(p \cos \chi) \hat{\bm{y}}, 
\end{equation}

where $p$ is the polarization fraction and $\chi$ is the polarization angle. We assumed a constant polarization fraction of $p = 0.1$, as in \citet{2017ApJ...835..137W}. The $\chi$ is derived from Stokes parameters; the relative Stokes parameters were calculated as follows: 

\begin{eqnarray}
\label{eq:q}
q =\int n \frac{B_{y}^{2}-B_{x}^{2}}{B^{2}} dx\\
\label{eq:u}
u = \int n \frac{2 B_{x} B_{y}}{B^{2}} dx, 
\end{eqnarray}
where $B$ is the magnetic field and $n$ is the density. The inferred polarization angle on the plane of the sky is given by the four-quadrant inverse tangent

\begin{equation}
 \label{eq:chi}
 \chi = \frac{1}{2} \arctan 2(u, q).  
\end{equation}

The polarization angle $\chi$ was measured clockwise from north.



\section{Results}
\label{sec:result}
We performed an analysis of the 27 simulation models. The physical parameters used in each model are listed in Table \ref{tab:Explored Parameter Space}. In the first column, we provide the model names used in the subsequent discussion. Section \ref{sec:Head-on collisions} shows the results of head-on collisions, whereas Section \ref{sec:off-center collisions} shows the result of an off-center collision. 

\subsection{Head-on collisions}
In this section, we present the results for the head-on ($b = 0$) collision cases. In all the head-on models, a single protostar (sink particle) was formed at the center of the shock-compressed layer. The accretion rates on the protostars depended on the collision parameters. 
In Section \ref{subsec:The general picture of head-on collision}, we describe the general picture of a head-on collision using the evolution of the slower pre-collision velocity of $\mathcal{M} = 1$, a weak magnetic field $B_{0} = 10~\mu G$, and $\theta = 0^{\circ}$ model {\tt b0-M1-By10}. In Section \ref{subsec:M=1, theta=0}$-$\ref{subsec:M=3, theta=0}, we show how variations in the collision parameters affect the formation of protostars. 


\label{sec:Head-on collisions}

\begin{figure*}
    \begin{center}
    \includegraphics[clip,height=0.5\hsize]{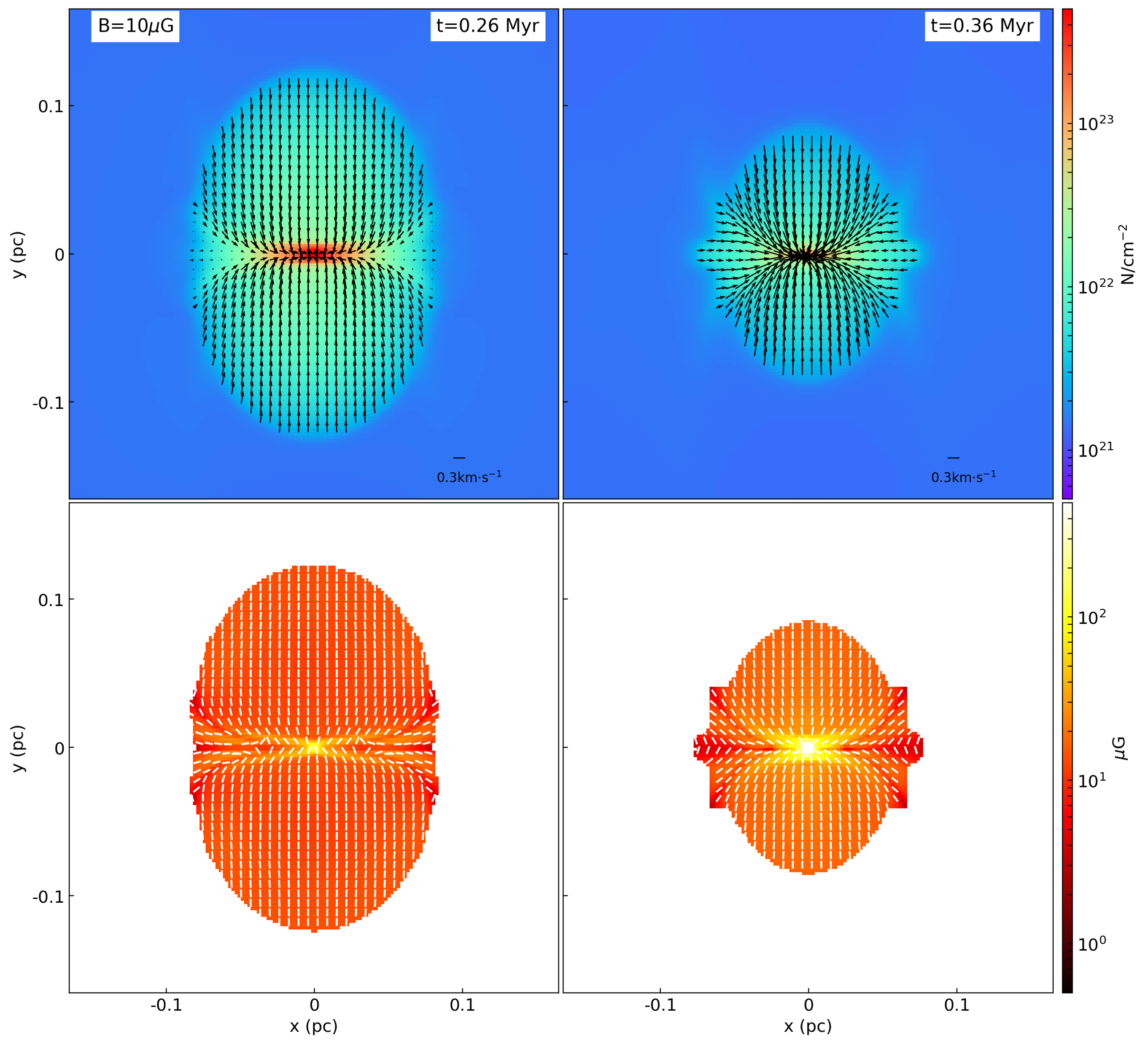}
     
    \vskip5pt  
    \caption{Column density, magnetic field, and velocity distributions of the model {\tt b0-M1-By10}. \textit{First row}: column density maps. The black vectors indicate the mass-weighted velocity field. Velocity arrows scale linearly with respect to the reference velocity shown in the bottom right of the panel and corresponding to $0.3 ~\rm km\cdot s^{-1}$. \textit{Second row}: mass-weighted magnetic field strength of cloud material are shown directly below their corresponding column-density maps. White pseudovectors indicate the normalized plane-of-sky magnetic polarization field, $\bm{p}$. 
\label{fig:b0-M1-By10}}
    
\end{center}
\end{figure*}

\subsubsection{Head-on collision overview}
\label{subsec:The general picture of head-on collision}
Figure \ref{fig:b0-M1-By10} shows the column density, velocity, and magnetic field distributions for model {\tt b0-M1-By10}. The left column shows a snapshot when the sink particle formed. The right column is 0.1 Myr after sink formation. In the first row, we show the column density map with mass-weighted velocity vectors in the $x-y$ plane. During the collision process, a shock-compressed layer parallel to the $x$-axis direction formed at the interface between the two cores. At this collision front, the direction of gas motion was changed parallel to the $x$-axis, and some gas flowed into the center. The material continued to accrete into the center, forming a high-density point where a sink particle formed. Table \ref{tab:simulation result} lists the sink particle mass and star formation efficiency (SFE). As in \citet{2007MNRAS.378..507K}, we assumed that the Class 0 phase lasts for $\sim 0.1$ Myr and that the final mass of the protostar is equal to the mass of the sink particle 0.1 Myr after formation. For the model {\tt b0-M1-By10}, the final mass was $4.9~M_{\odot}$, and SFE was $67$ \%. Most of the core material accreted onto the central particle during collision. 


The second row of Figure \ref{fig:b0-M1-By10} shows the mass-weighted plane-of-sky magnetic field strength of the cloud material. White pseudovectors indicate the normalized magnetic polarization field, $\bm{p}$ derived using Equations \ref{eq:pseudo-vector}–\ref{eq:chi}. In the central area, we can see the hourglass shape of the magnetic field with a stronger central field. Accreting gas dragged the magnetic field lines inward in densely contracted regions, increasing the field strength. However, at the edge, the magnetic field bent nearly along the $x$-direction, and the field strength decreased. 

\subsubsection{$\mathcal{M}=1$ and $\theta=0^{\circ}$}
\label{subsec:M=1, theta=0}

\begin{figure*}
  
    \begin{tabular}{c}

      \begin{minipage}{0.5\hsize}
        \begin{center}
          \includegraphics[clip, width=8.0cm]{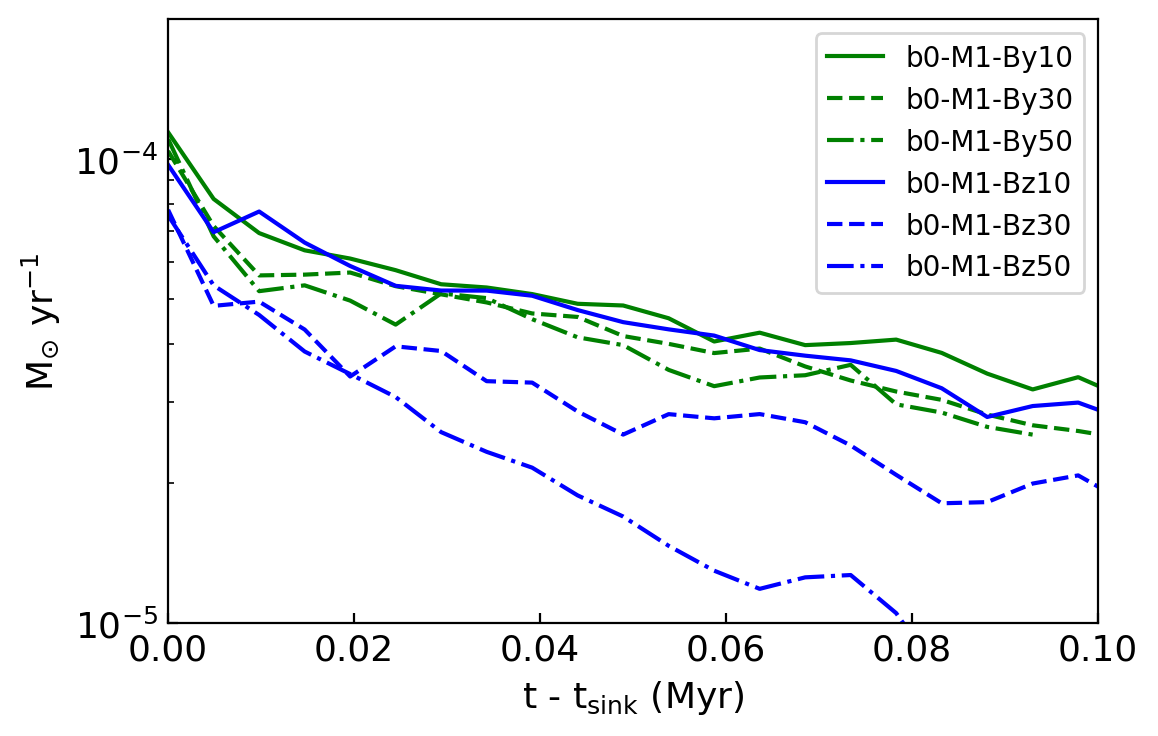}
        \end{center}
      \end{minipage}

      \begin{minipage}{0.5\hsize}
        \begin{center}
          \includegraphics[clip, width=8.0cm]{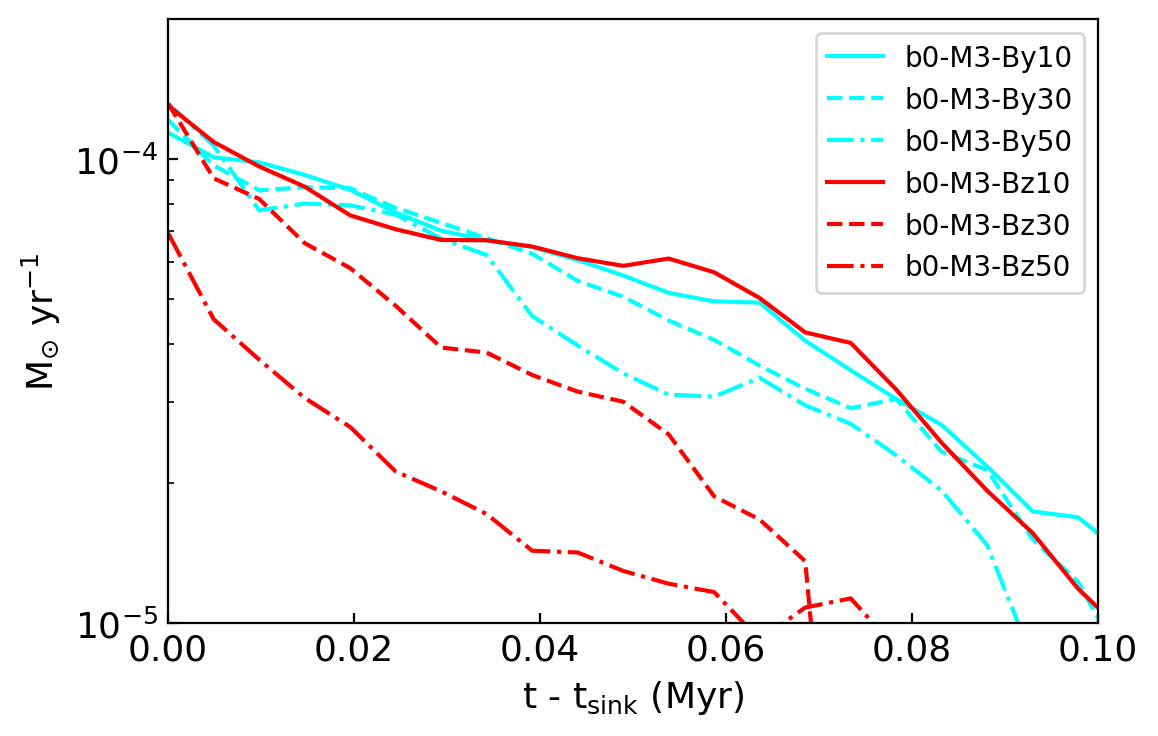}
        \end{center}
      \end{minipage}

       \end{tabular}

    \vskip5pt  
    \caption{Temporal evolution of mass accretion rate of sink particles for head-on $(b = 0)$ cases. \textit{Left panel}: $\mathcal{M} = 1$ models. \textit{Right panel}: $\mathcal{M} = 3$ models.
\label{fig:b0-accretion_rate}}

\end{figure*}

\begin{figure*}
    \begin{center}
    \includegraphics[clip,height=0.3\hsize]{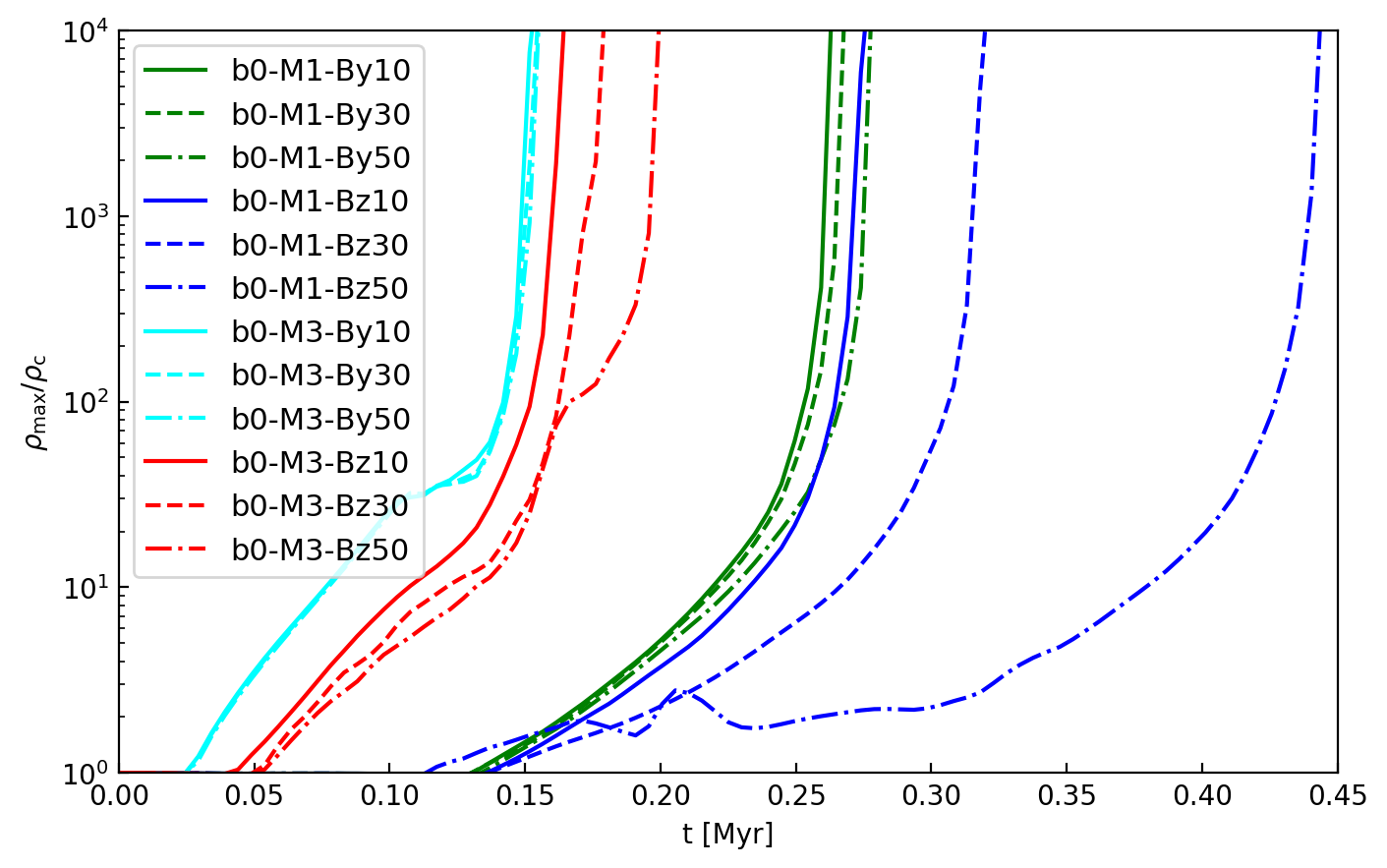}
     
    \vskip5pt  
    \caption{Temporal evolution of the maximum density normalized to the initial central core density for head-on $(b = 0)$ cases.
\label{fig:rho_max}}
    
\end{center}
\end{figure*}

The $\mathcal{M} = 1$ and $\theta = 0^{\circ}$ cases were explored in models {\tt b0-M1-By10}, {\tt b0-M1-By30}, and {\tt b0-M1-By50}. Figure \ref{fig:b0-accretion_rate} shows the temporal evolution of the mass accretion rate of sink particles. This figure and Table \ref{tab:simulation result} show that comparing the three models, the higher the magnetic field strength, the lower the accretion rate. Models {\tt b0-M1-By10}, {\tt b0-M1-By30}, and {\tt b0-M1-By50} had SFEs values of 67\%, 61\%, and 56\%, respectively. The average mass accretion rates within 0.1 Myr were $10^{-5}-10^{-4}~M_{\odot}~\rm yr^{-1}$, which is a typical value for Class 0 protostars \citep{1996A&A...311..858B}.

\subsubsection{$\mathcal{M}=1$ and $\theta=90^{\circ}$}
\label{subsec:M=1, theta=90}
The $\mathcal{M} = 1$ and $\theta = 90^{\circ}$ cases were explored in models {\tt b0-M1-Bz10}, {\tt b0-M1-Bz30}, and {\tt b0-M1-Bz50}. Figure \ref{fig:rho_max} shows the temporal evolution of maximum density in each model. Comparing models {\tt b0-M1-Bz10}, {\tt b0-M1-Bz30}, and {\tt b0-M1-Bz50}, shown by blue lines, the higher the magnetic field strength, the slower the rate of increase in density. Therefore, as shown in Table \ref{tab:simulation result}, the sink formation time $t_{\rm sink}$ also occurred later in the stronger magnetic field model. 

The left panel in Figure \ref{fig:b0-accretion_rate} and Table \ref{tab:simulation result} show that the higher the magnetic field strength, the lower the accretion rate. For the weak magnetic field model {\tt b0-M1-Bz10}, the SFE was 70\%, whereas for the strong magnetic field model {\tt b0-M1-Bz50}, it was less than half (34\%). The difference in accretion rate among these three models was larger than that of the $\mathcal{M} = 1$ and $\theta = 0^{\circ}$ models. The $\theta = 0^{\circ}$ models {\tt b0-M1-By10} and {\tt b0-M1-By50} had SFEs of 67\% and 56\%, respectively, whereas the $\theta = 90^{\circ}$ model {\tt b0-M1-Bz50} had less than half the SFE of the $\theta = 90^{\circ}$ model {\tt b0-M1-Bz10} because of the increase in the magnetic field strength.

\subsubsection{$\mathcal{M}=3$ and $\theta=0^{\circ}$}
\label{subsec:M=3, theta=0}
Models {\tt b0-M3-By10}, {\tt b0-M3-By30}, and {\tt b0-M3-By50} explored the case of a faster pre-collision velocity $\mathcal{M} = 3$ and $\theta = 0^{\circ}$. Figure \ref{fig:rho_max} shows that for these models, the density increased more rapidly than in the slower velocity models ($\mathcal{M} = 1$). This is because of the stronger compression over a short period of time. 
In addition, Table \ref{tab:simulation result} shows that the mass accretion rates and SFEs were higher than those of the slower velocity models. While, for the faster velocity models {\tt b0-M3-By10}, {\tt b0-M3-By30}, and {\tt b0-M3-By50}, SFEs were 85\%, 80\%, and 71\%, respectively. For the slower velocity models {\tt b0-M1-By10}, {\tt b0-M1-By30}, and {\tt b0-M1-By50}, SFEs were 67\%, 61\%, and 56\%, respectively.


\subsubsection{$\mathcal{M}=3$ and $\theta=90^{\circ}$}
\label{subsec:M=3, theta=90}
Models {\tt b0-M3-Bz10}, {\tt b0-M3-Bz30}, and {\tt b0-M3-Bz50} explored the case of a faster pre-collision velocity of $\mathcal{M} = 3$ and $\theta = 90^{\circ}$. The density increased faster than that in the slower velocity models. Figure \ref{fig:b0-accretion_rate} shows that the difference in accretion rate among models was larger than that of the $\mathcal{M} = 3$ and $\theta = 0^{\circ}$ models. For models {\tt b0-M3-By10}, {\tt b0-M3-By30}, and {\tt b0-M3-By50}, SFEs were 85\%, 80\%, and 71\%, respectively. However, for the $\mathcal{M} = 3$ and $\theta = 90^{\circ}$ models {\tt b0-M3-Bz10}, {\tt b0-M3-Bz30}, and {\tt b0-M3-Bz50}, SFEs were 85\%, 54\%, and 35\%, respectively.

\begin{deluxetable*}{lccccl}
\tablecaption{Simulation results}
\tablewidth{0pt}
\tablehead{
Model name & $t_{\rm sink}$ $^{\rm a}$ & $\langle \dot{M}\rangle$ $^{\rm b}$ & $M_{*}$ $^{\rm c}$& SFE $^{\rm d}$& Results\\
           & (Myr) & ($M_{\odot} \cdot \rm yr^{-1}$) &  ($M_{\odot}$) & ($\%$)
}
\startdata
{\tt b0-M1-By10}&  0.26  & $5.3\times10^{-5}$ & 4.9 &    67    &  single star        \\  
{\tt b0-M1-By30}&  0.26  & $4.7\times10^{-5}$ & 4.4 &    61    &  single star        \\
{\tt b0-M1-By50}&  0.27  & $4.4\times10^{-5}$ & 4.1 &    56    &  single star        \\
\hline
{\tt b0-M1-Bz10}&  0.27  & $4.9\times10^{-5}$ & 5.1 &    70    &  single star       \\  
{\tt b0-M1-Bz30}&  0.32  & $3.3\times10^{-5}$ & 3.6 &    50    &  single star        \\
{\tt b0-M1-Bz50}&  0.44  & $2.4\times10^{-5}$ & 2.5 &    34    &  single star        \\
\hline
{\tt b0-M3-By10}&  0.15  & $6.0\times10^{-5}$ & 6.2 &    85    &  single star        \\  
{\tt b0-M3-By30}&  0.15  & $5.7\times10^{-5}$ & 5.8 &    80    &  single star        \\
{\tt b0-M3-By50}&  0.15  & $5.1\times10^{-5}$ & 5.2 &    71    &  single star        \\
\hline
{\tt b0-M3-Bz10}&  0.16  & $6.1\times10^{-5}$ & 6.2 &    85    &  single star       \\  
{\tt b0-M3-Bz30}&  0.18  & $3.7\times10^{-5}$ & 4.0 &    54    &  single star        \\
{\tt b0-M3-Bz50}&  0.20  & $2.0\times10^{-5}$ & 2.5 &    35    &  single star        \\
\hline
\hline
\multirow{2}{*}{\tt b1-M1-By10}&  0.31  & $1.4\times10^{-5}$ & 1.4 &    \multirow{2}{*}{38}   &  \multirow{2}{*}{binary}        \\  
                &  0.31  & $1.4\times10^{-5}$ & 1.4 &     &         \\  
{\tt b1-M1-By30}&  0.36  & $2.5\times10^{-5}$ & 2.4 &   33      &  single star        \\
{\tt b1-M1-By50}&  0.53  & $2.5\times10^{-5}$ & 2.4 &   33      &  single star         \\
\hline
\multirow{2}{*}{\tt b1-M1-Bz10}&  0.30  & $1.4\times10^{-5}$ & 1.4 &    \multirow{2}{*}{39}   &  \multirow{2}{*}{binary}        \\  
                &  0.30  & $1.4\times10^{-5}$ & 1.4 &     &         \\  
\multirow{2}{*}{\tt b1-M1-Bz30}&  0.40  & $1.7\times10^{-5}$ & 1.8 &    \multirow{2}{*}{49}   &  \multirow{2}{*}{binary}        \\  
                &  0.40  & $1.7\times10^{-5}$ & 1.8 &     &         \\  
{\tt b1-M1-Bz50}&  ---  & --- & 0  & 0    &  merge        \\
\hline
\multirow{2}{*}{\tt b1-M1-Bob10}&  0.30  & $1.3\times10^{-5}$ & 1.4 &    \multirow{2}{*}{36}   &  \multirow{2}{*}{binary}        \\  
                &  0.30  & $1.3\times10^{-5}$ & 1.4 &     &         \\  
\multirow{2}{*}{\tt b1-M1-Bob30}&  0.46  & $1.7\times10^{-5}$ & 1.8 &    \multirow{2}{*}{45}   &  \multirow{2}{*}{binary}        \\  
                &  0.46  & $1.6\times10^{-5}$ & 1.8 &     &         \\  
{\tt b1-M1-Bob50}&  ---  & --- & 0  & 0    &  merge        \\
\hline
{\tt b1-M3-By10}&  ---  & --- & 0 &  0   &  destruction        \\  
{\tt b1-M3-By30}&  ---  & --- & 0 &  0   &  destruction        \\
{\tt b1-M3-By50}&  ---  & --- & 0 &  0   &  destruction        \\
\hline
{\tt b1-M3-Bz10}&  ---  & --- & 0 &  0  &  destruction        \\  
{\tt b1-M3-Bz30}&  ---  & --- & 0 &  0  &  destruction        \\
{\tt b1-M3-Bz50}&  ---  & --- & 0 &  0  &  destruction        \\
\enddata
\tablenotetext{\tiny \rm{a}}{Time interval from collision start until sink particles formed.}
\tablenotetext{\tiny \rm{b}}{Mean mass accretion rate within 0.1 Myr after sink formation.}
\tablenotetext{\tiny \rm{c}}{Sink particle mass 0.1 Myr after sink formation.}
\tablenotetext{\tiny \rm{d}}{Core-to-sink efficiency 0.1 Myr after sink formation.}
\label{tab:simulation result}
\end{deluxetable*}

\subsection{Off-center collisions}
\label{sec:off-center collisions}
In this section, we present the results for the off-center ($b = 1$) collision cases. 
As shown below, the number of protostars and accretion rates were highly dependent on the collision parameters.

\subsection{$\mathcal{M}=1$, and $\theta=0^{\circ}$}
\label{sec:off-center perpendicular}
Figure \ref{fig:b1-M1-By10} shows snapshots of the weak magnetic-field model {\tt b1-M1-By10}. 
Initially, at $t\sim0.21 ~\rm Myr$, a shocked slab layer formed between the two cores. The gas and magnetic fields were bent parallel to the slab. The gas rotated clockwise because of the orbital angular momentum in the collision process at $b = 1$. Then, at $t = 0.31 ~\rm Myr$, the shocked layer fragmented, and two sink particles were created; that is, binary forms. The separation between the particles was $\sim 0.01$ pc and the particles were gravitationally bound. The gas was accreting onto the particles with an increasing specific angular momentum. The circumbinary disc developed, and the two protostars grew to $\sim 1~M_{\odot}$. The average mass accretion rates were $ \sim 1.4\times10^{-5}~ M_{\odot}~\rm yr^{-1}$ and the SFE was 38\%. The magnetic fields were aligned with the gas inflow.

In the intermediate magnetic field model {\tt b1-M1-By30}, a single star is formed, rather than a binary system. Figure \ref{fig:b1-M1-By30} presents snapshots of the intermediate magnetic field model {\tt b1-M1-By30}. Initially, a shock layer was formed. Owing to the orbital angular momentum, clockwise rotational movement began. Subsequently, at $t = 0.36$ Myr, the shock-compressed layer contracted gravitationally and a single sink particle was formed. The gas was accreting onto the particles with an increasing specific angular momentum. At $t = 0.46$ Myr a two-armed spiral developed. The average mass accretion rate was $ 2.4\times10^{-5} ~M_{\odot}~\rm yr^{-1}$ and SFE was 33\%. 

Additionally, in the strong magnetic field model {\tt b1-M1-Bz50}, a single star formed with a two-armed spiral. The SFE was almost the same as that of the intermediate magnetic field model {\tt b1-M1-By30}.

\subsubsection{$\mathcal{M}=1$, and $\theta=90^{\circ}$}
\label{subsec:b=1 M=1 theta=90}

\begin{figure*}
    \begin{center}
    \includegraphics[clip,height=0.5\hsize]{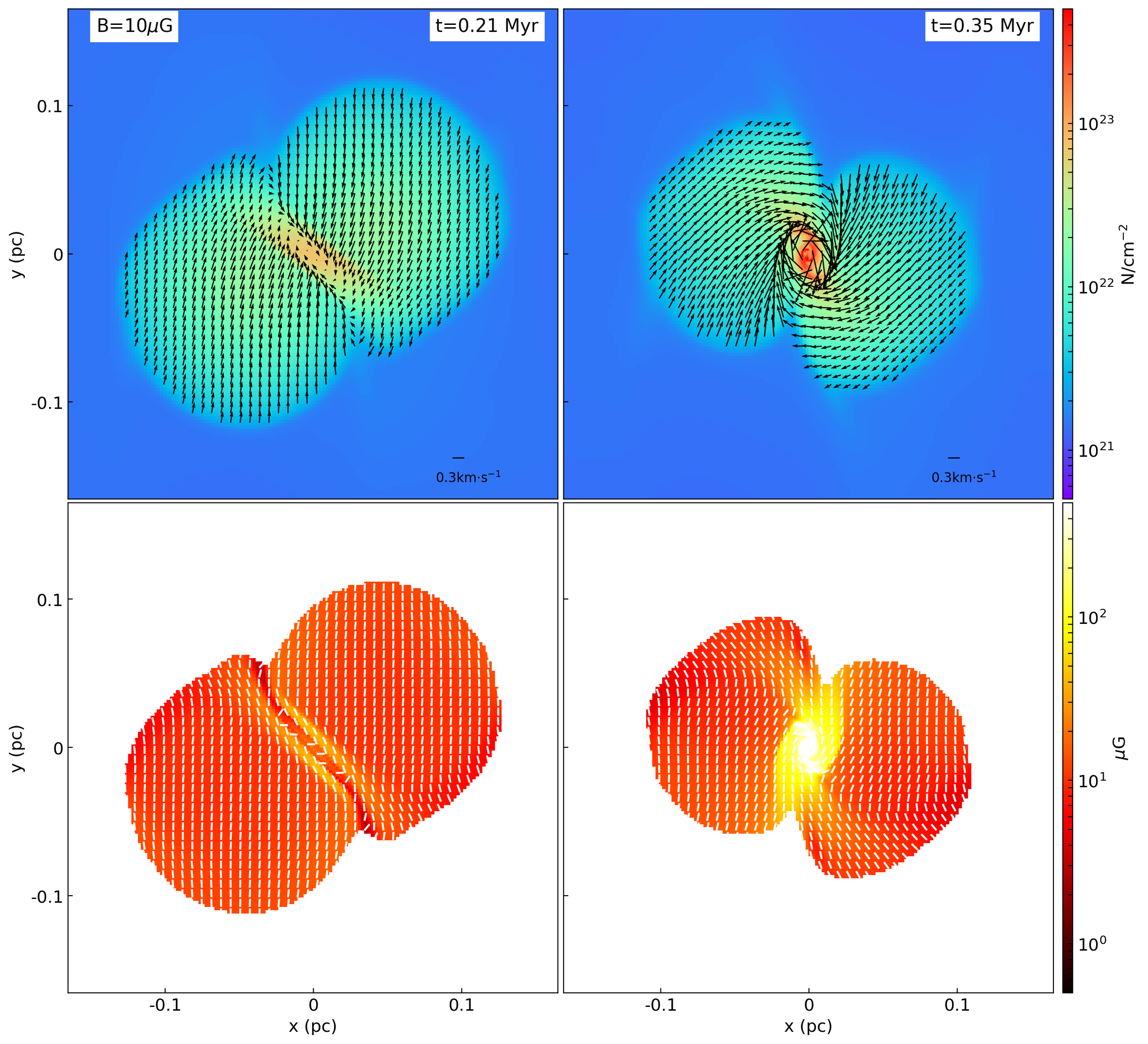}
     
    \vskip5pt  
    \caption{As in Figure \ref{fig:b0-M1-By10} for the model {\tt b1-M1-Bz10}. Snapshots at 0.21, 0.3, and 0.4 Myr are shown. 
\label{fig:b1-M1-By10}}
    
\end{center}
\end{figure*}

For the weak magnetic field model {\tt b1-M1-Bz10}, binary formed with a mean accretion rate of $\dot{M} = 1.4\times 10^{-5}~M_{\odot}~\rm yr^{-1}$ and SFE of 39\%. Figure \ref{fig:b1-M1-By10} shows the snapshots. The separation between the particles was larger than that of the $\theta = 0^{\circ}$ model {\tt b1-M1-By10} and $\sim 0.04$ pc.

\begin{figure*}
    \begin{center}
    \includegraphics[clip,height=0.3\hsize]{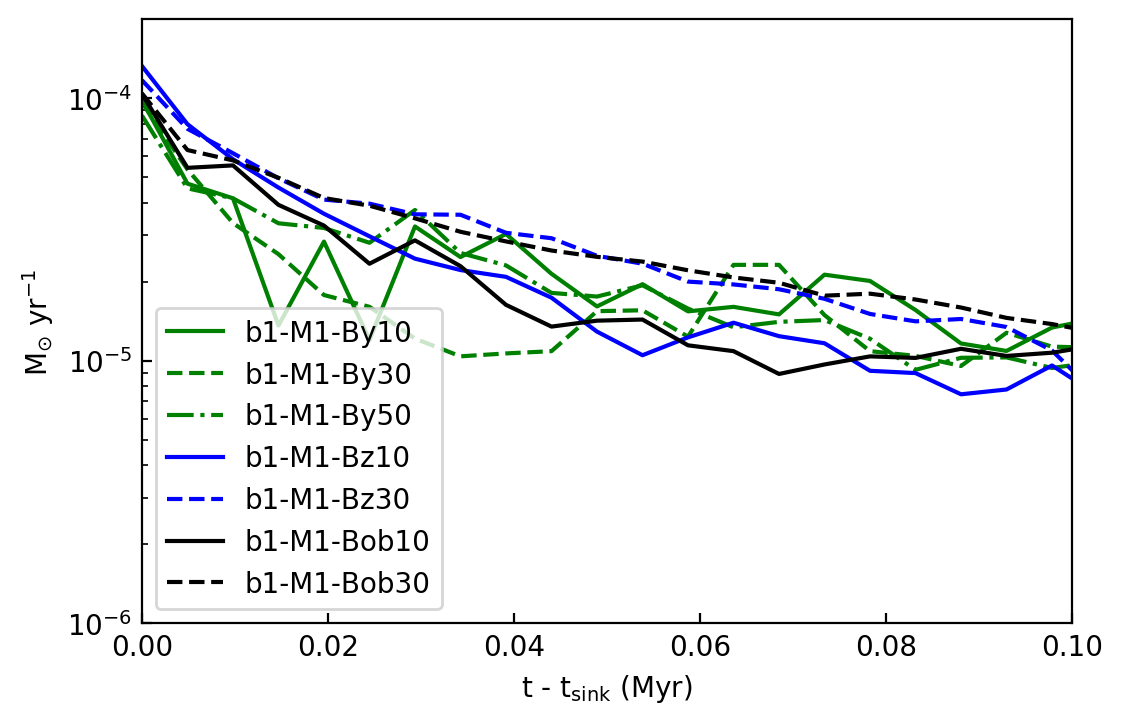}
     
    \vskip5pt  
    \caption{Temporal evolution of the mass accretion rate of sink particles for off-center $(b = 1)$ cases. In cases where the binary is formed, the total accretion rates of two sinks are shown.
\label{fig:b1-accretion_rate}}
    
\end{center}
\end{figure*}

For the intermediate magnetic field model {\tt b1-M1-Bz30}, the result differed from that of the $\theta = 0^{\circ}$ model {\tt b1-M1-By30}. In this model, a binary system formed. Figure \ref{fig:b1-accretion_rate} shows the evolution of the accretion rates. Most of the time, the accretion rate was higher than that in the weak magnetic field model {\tt b1-M1-Bz10}. For model {\tt b1-M1-Bz30}, SFE was 49\%, whereas for model {\tt b1-M1-Bz10}, SFE was 39\%.

\begin{figure*}
    \begin{center}
    \includegraphics[clip,height=0.3\hsize]{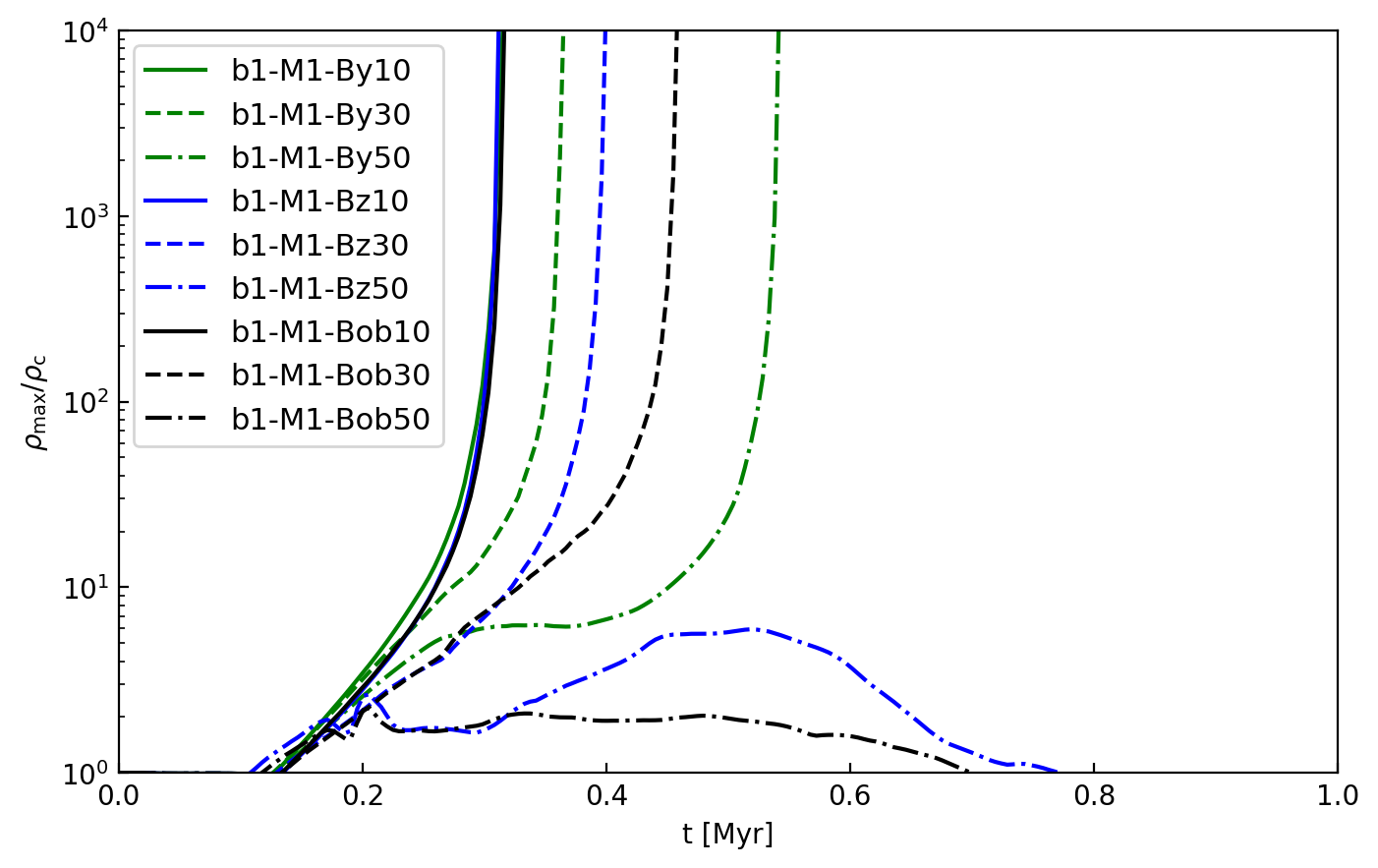}
     
    \vskip5pt  
    \caption{As in Figure \ref{fig:rho_max} for off-center $(b = 1)$ cases.
\label{fig:rho_max_b1}}
    
\end{center}
\end{figure*}

Figure \ref{b1-M1-Bz50} shows snapshots of the strong magnetic model {\tt b1-M1-Bz50}. The density increased in the shocked layer at approximately $t = 0.21 ~\rm Myr$. However, unlike the weak and intermediate magnetic models {\tt b1-M1-Bz10} and {\tt b1-M1-Bz30}, no structure that was dense enough for protostar creation was formed (see also Figure \ref{fig:rho_max_b1}). The two cores rotated slowly and merged gradually to form a single longitudinal structure. Following the evolution to $t\sim 1.0$ Myr ($\sim 10~t_{\rm ff})$, we found that the structure remained almost the same without collapsing.

\subsubsection{$\mathcal{M}=1$, and $\theta=45^{\circ}$}
\label{subsec:b=1 M=1 theta=45}

For the weak magnetic field model {\tt b1-M1-Bob10}, the binary formed with mean accretion rates of $\dot{M} = 1.3\times 10^{-5}~M_{\odot}~\rm yr^{-1}$ and SFE of 36\%. This accretion rate is almost the same as that of {\tt b1-M1-By10} and {\tt b1-M1-Bz10}. For the intermediate magnetic field model {\tt b1-M1-Bob30}, a binary system formed with a mean accretion rate of $\dot{M} = 1.7\times 10^{-5}~M_{\odot}~\rm yr^{-1}$ and SFE of 45\%. For the strong magnetic field model {\tt b1-M1-Bob50}, similar to the $\theta = 90^{\circ}$ model {\tt b1-M1-Bz50}, star formation was not triggered. The two cores merged gradually without collapse.

\subsubsection{$\mathcal{M}=3$}
\begin{figure*}
    \begin{center}
    \includegraphics[clip,height=0.5\hsize]{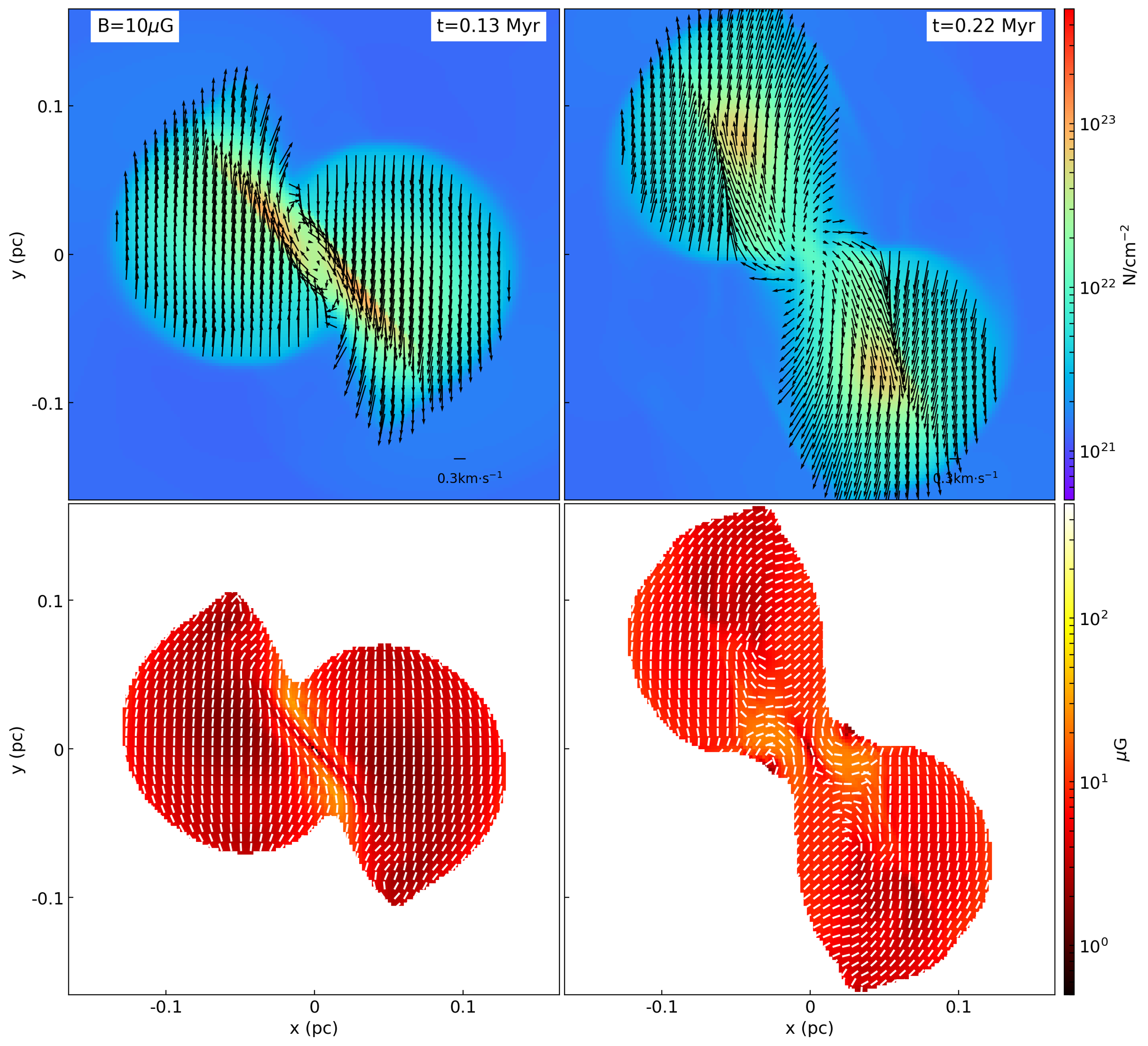}
     
    \vskip5pt  
    \caption{As in Figure \ref{fig:b0-M1-By10} for the model {\tt b1-M3-Bz10}. Snapshots at 0.13 and 0.22 Myr are shown. 
\label{fig:b1-M3-Bz10}}
    
\end{center}
\end{figure*}

In all $b = 1$ and $\mathcal{M} = 3$ models, star formation was not triggered. Figure \ref{fig:b1-M3-Bz10} shows the temporal evolution of model {\tt b1-M3-Bz10}. For this collision, no significant shock-compressed layer was formed, and the two cores passed each other. As the cores progressed, the Kelvin-Helmholtz (K-H) instability between the cores and the ambient gas was excited, leading to ablation of the cores \citep{2004ApJ...615..586M}. At $t\sim 0.5$ Myr, the two cores touched the boundary of the simulation box and were mixed with the ambient gas. The results are similar for $\mathcal{M} = 3$. No significant shock-compressed layer was formed and star formation was not triggered during the collision process. These results are consistent with those of \citet{2007MNRAS.378..507K}, who investigated clump collisions using hydrodynamic simulations. \citet{2007MNRAS.378..507K} showed that large-$b$ and high-$M$ collisions reduced clump collisions, and no shock-compressed layer was formed.

\begin{figure*}
    \begin{center}
    \includegraphics[clip,height=0.5\hsize]{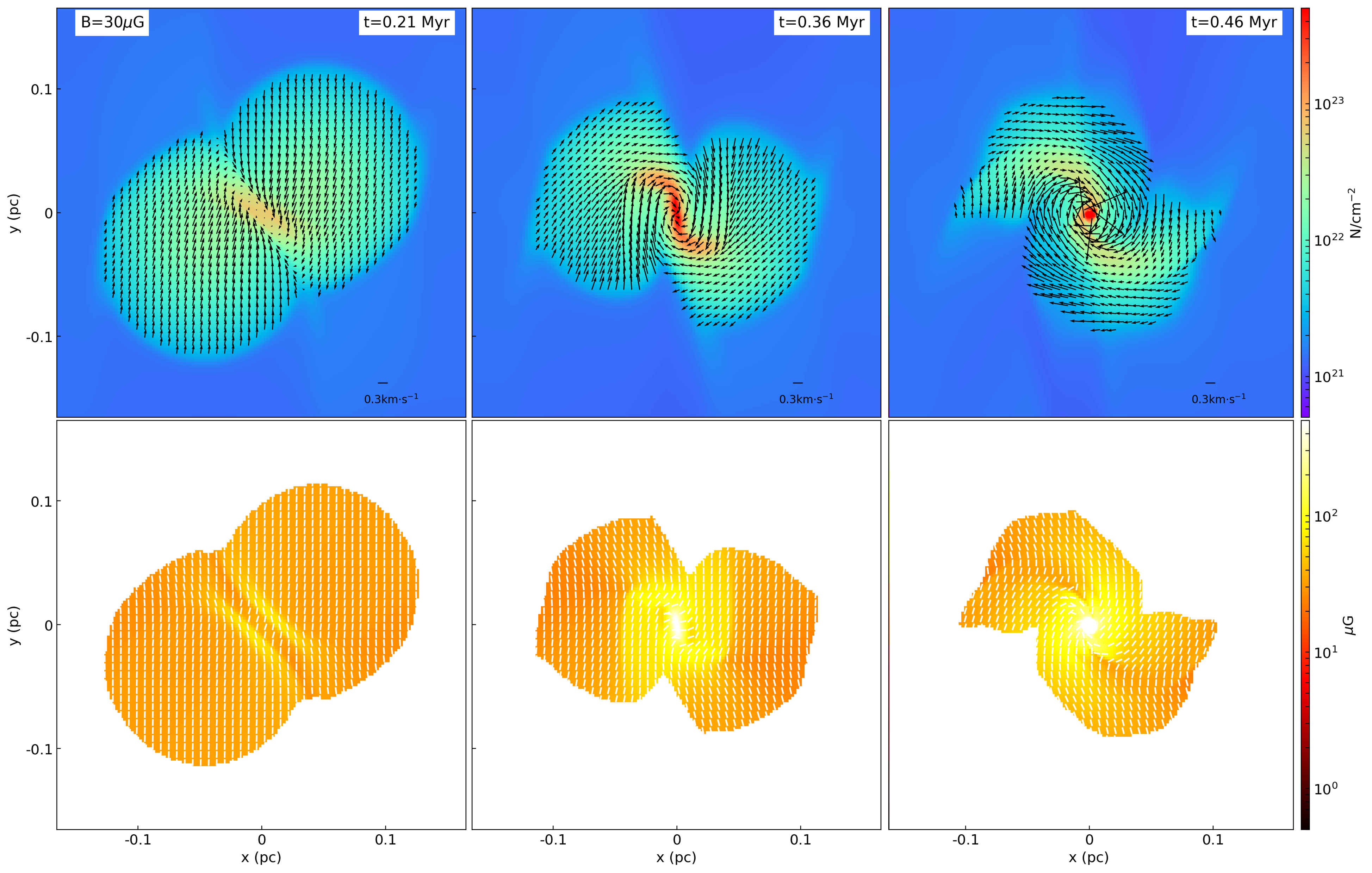}
     
    \vskip5pt  
    \caption{As in Figure \ref{fig:b0-M1-By10} for the model {\tt b1-M1-By30}. Snapshots at 0.21, 0.36, and 0.46 Myr are shown. 
\label{fig:b1-M1-By30}}
    
\end{center}
\end{figure*}

\begin{figure*}
    \begin{center}
    \includegraphics[clip,height=0.5\hsize]{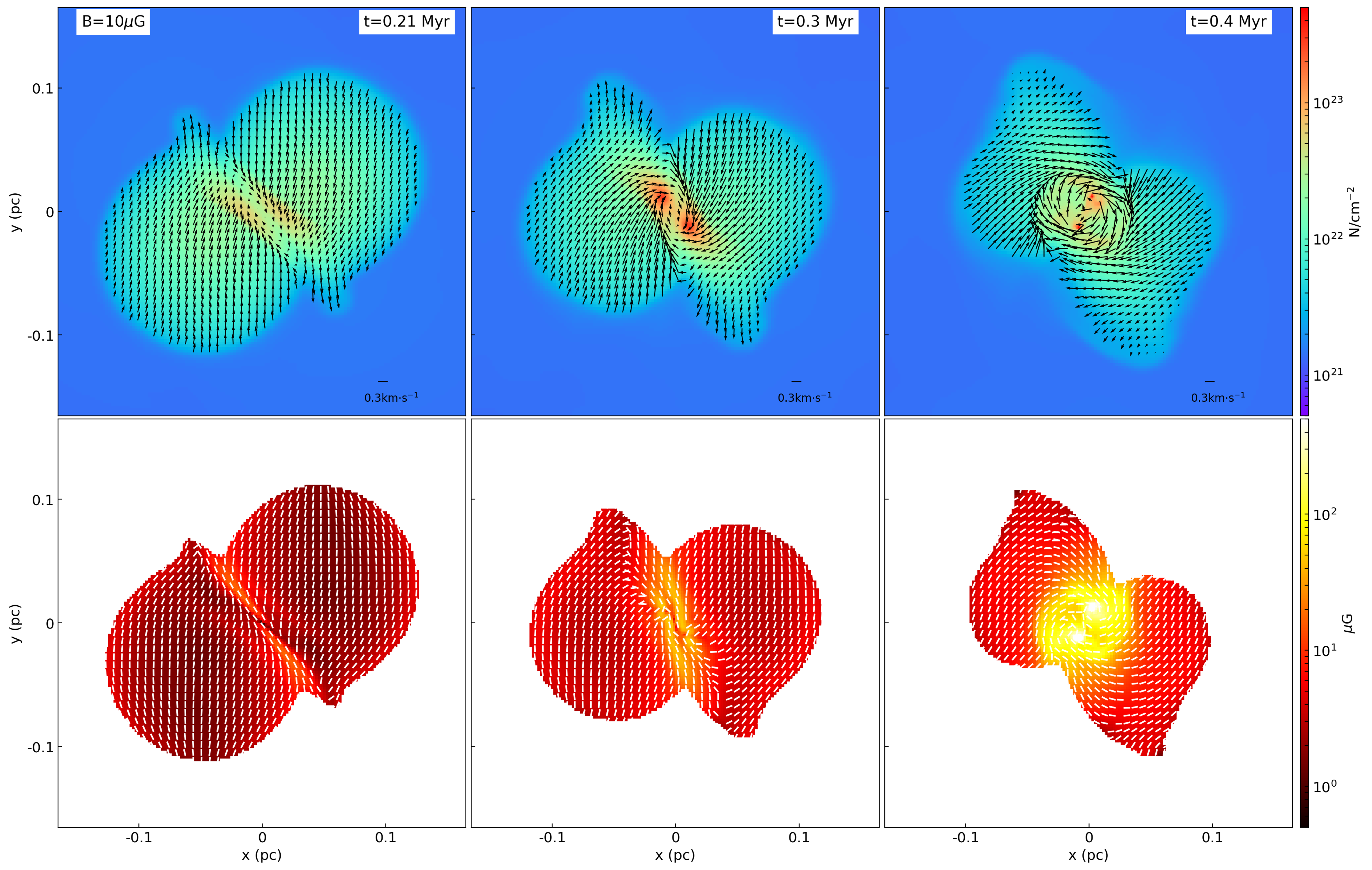}
     
    \vskip5pt  
    \caption{As in Figure \ref{fig:b0-M1-By10} for the model {\tt b1-M1-Bz10}. Snapshots at 0.21, 0.3, and 0.4 Myr are shown. 
\label{fig:b1-M1-Bz10}}
    
\end{center}
\end{figure*}

\begin{figure*}
    \begin{center}
    \includegraphics[clip,height=0.5\hsize]{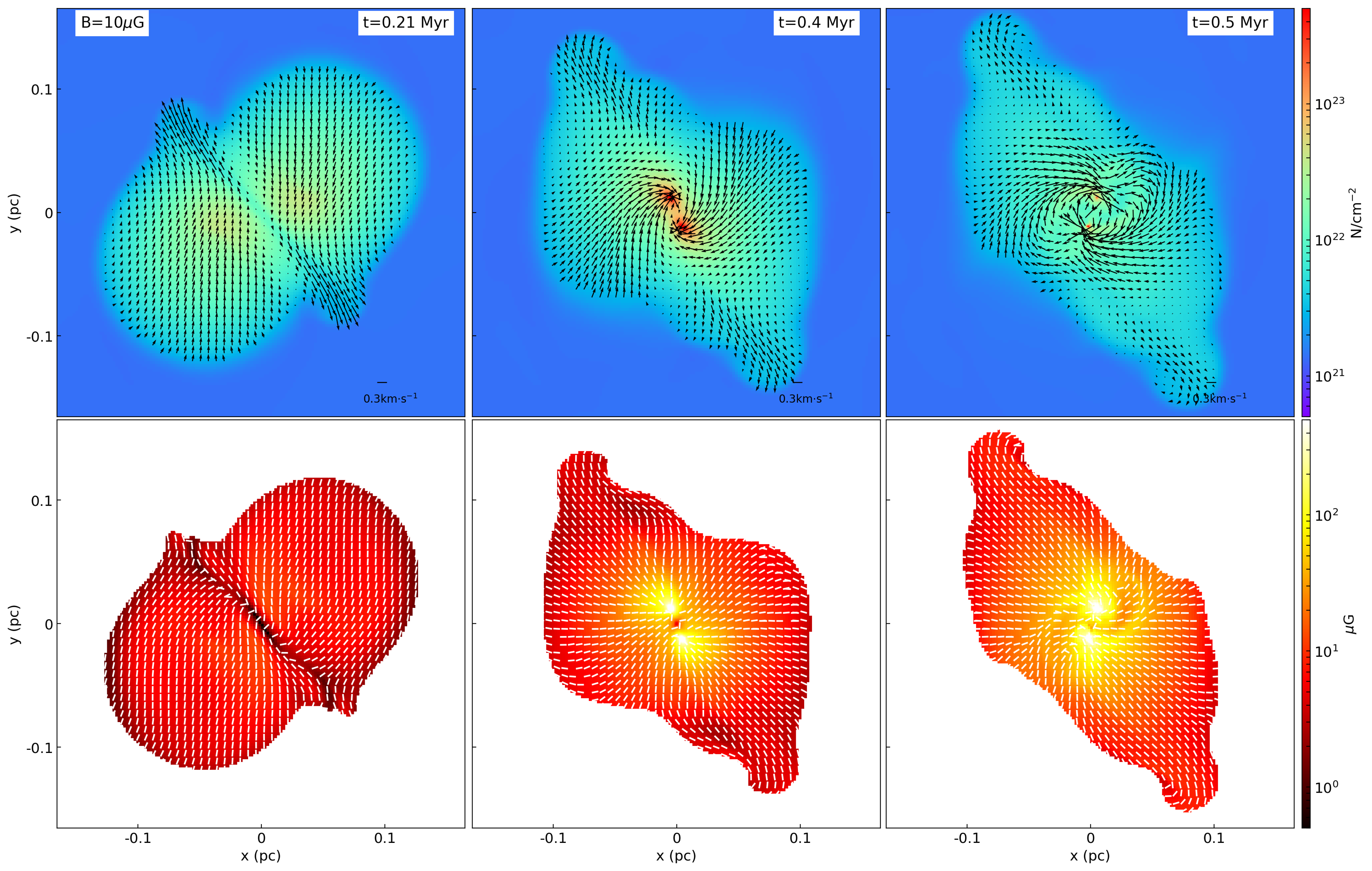}
     
    \vskip5pt  
    \caption{As in Figure \ref{fig:b0-M1-By10} for the model {\tt b1-M1-Bz30}. Snapshots at 0.21, 0.4, and 0.5 Myr are shown. 
\label{fig:b1-M1-Bz30}}
    
\end{center}
\end{figure*}

\begin{figure*}
    \begin{center}
    \includegraphics[clip,height=0.5\hsize]{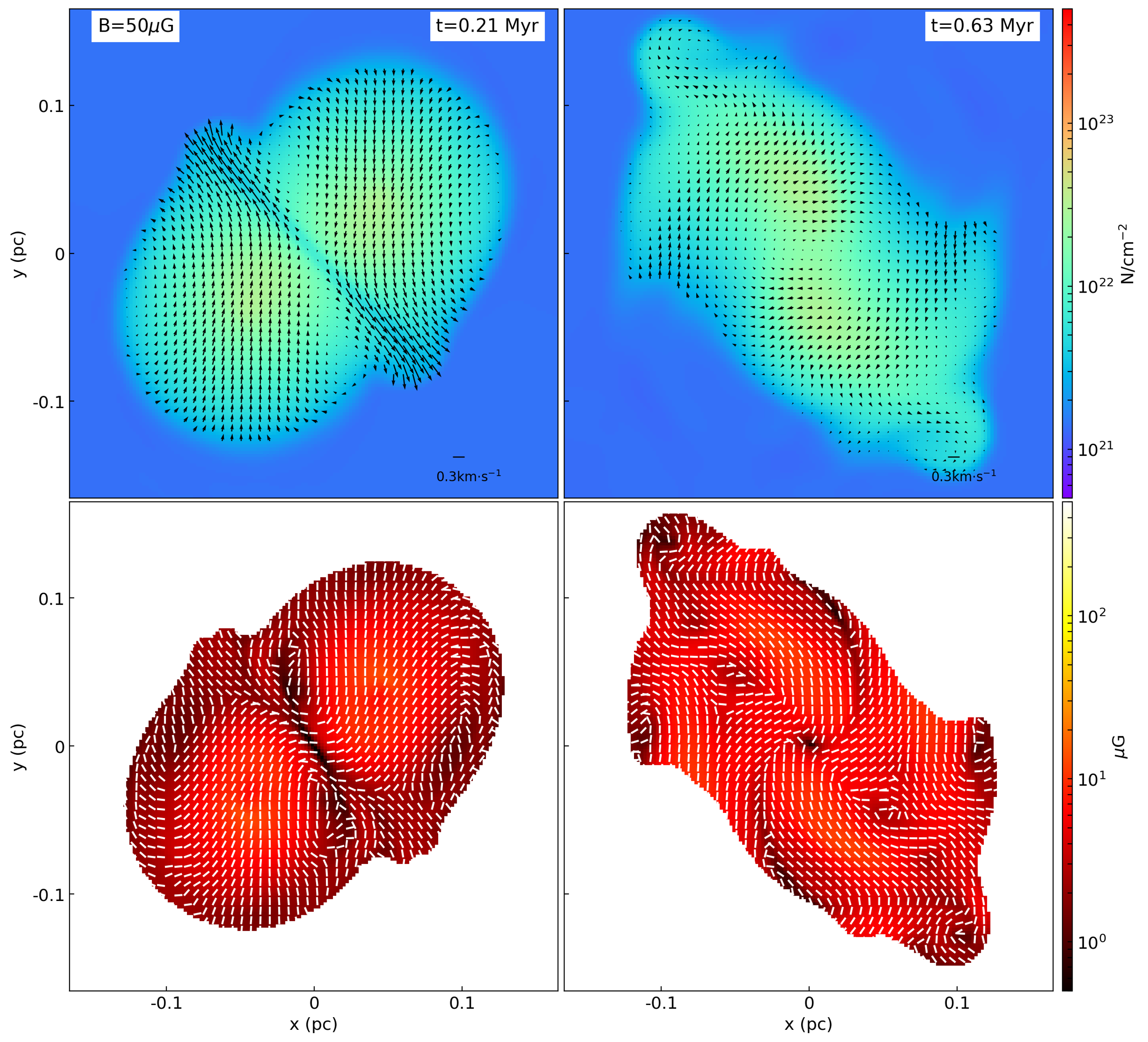}
     
    \vskip5pt  
    \caption{As in Figure \ref{fig:b0-M1-By10} for the model {\tt b1-M1-Bz50}. Snapshots at 0.21 and 0.63 Myr are shown. 
\label{b1-M1-Bz50}}
    
\end{center}
\end{figure*}

\section{Discussion}
\label{sec:discussion}
We performed a set of MHD simulations aimed at following the dense core collision process and the role of magnetic fields. As shown in Section \ref {sec:result}, there is some dependence of the structures formed during the collision process on the magnetic field parameter.
In this section, we discuss the role of the magnetic field. Section \ref{sec:discussion-head-on} covers head-on collision cases and Section \ref{sec:discussion-off-center} covers off-center cases. In Section \ref{sec:Mixing of gas}, we supplementarily discuss the gas mixing.

\subsection{Discussion of head-on collision}
\label{sec:discussion-head-on}
For the $\mathcal{M} = 1$ and $\theta = 90^{\circ}$ models, as the magnetic field strength $B_{0}$ increased, there were two main trends. First, the rate of increase in density decreased; therefore, the sink particle formation time $t_{0}$ became later. Second, the accretion rate on the protostar decreased, and therefore the protostar mass decreased. From these results, we conclude that the magnetic field hinders the formation of high-density regions and reduces accretion.

On the other hand, for the $\mathcal{M} = 1$ and $\theta = 0^{\circ}$ models, the dependence of the accretion rate on the field strength $B_{0}$ was lower. 
As magnetic pressure acts in directions perpendicular to the field lines, models in which the initial magnetic fields were parallel to the pre-collision velocity ($\theta = 0^{\circ}$) resulted in less inhibited flow and higher accretion rates. For cases where the initial magnetic fields were perpendicular to the pre-collision velocity ($\theta = 90^{\circ}$), the magnetic pressure effectively prevents gas flow. Therefore, in models with a higher magnetic field strength, the accretion rate was significantly lower. This trend was also true for $\mathcal{M} = 3$ model. Even in a simple process, such as a head-on collision, the accretion rate on the protostar strongly depends on the magnetic field direction.

\subsection{Discussion of off-center collision}
\label{sec:discussion-off-center}
\subsubsection{Effect of the initial magnetic field strength}
For the $\mathcal{M} = 1$ and $\theta = 0^{\circ}$ models, the number of protostars depended on the magnetic field strength $B_{0}$. Two protostars formed for the weak magnetic field model {\tt b1-M1-By10}. A single protostar formed for each of the intermediate and strong magnetic field models {\tt b1-M1-By30} and {\tt b1-M1-By50}.

\begin{figure*}
    \begin{center}
    \includegraphics[clip,height=0.3\hsize]{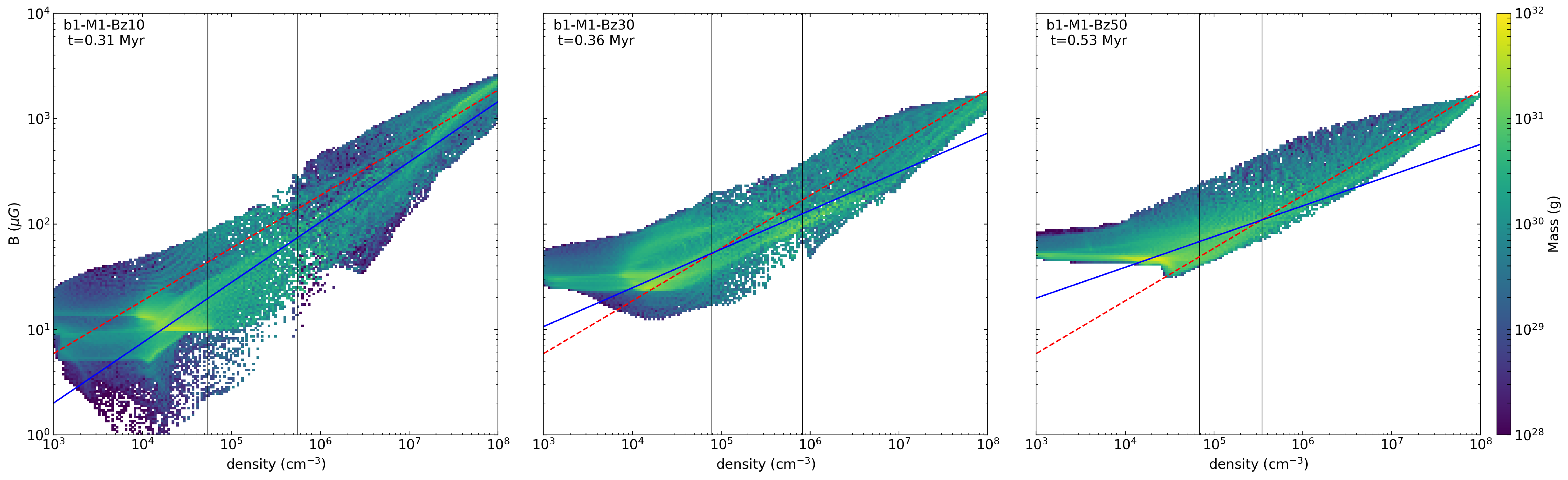}
     
    \vskip5pt  
    \caption{Phase plot of the magnetic field strength and gas density for models {\tt b0-M1-By10}, {\tt b0-M1-By30}, and {\tt b0-M1-By50} at $t_{\rm sink}$ for each model. The color bar displays the total gas mass at each point. Two black lines show that the total mass of the gas located to the right of these lines accounted for 50$\%$ and 25$\%$ of the total mass of the two cores. The red dashed line represents the Alfvén speed, $v_{\rm A} = c_{\rm s}$. Blue lines are the results of least-square fits with a power-law $B\propto n^{j}$. The derived $j$ for the models {\tt b0-M1-By10}, {\tt b0-M1-By30}, and {\tt b0-M1-By50} are 0.57, 0.37, and 0.29 respectively. 
\label{fig:n-B}}
    
\end{center}
\end{figure*}

The magnetic field energy significantly affects structures formed during the collision process. 
Figure \ref{fig:n-B} shows the magnetic field and density relationship for the {\tt b1-M1-By10}, {\tt b1-M1-By30}, and {\tt b1-M1-By50} models at $t = t_{\rm sink}$. We also drew constant Alfvén velocities of $1.0~c_{\rm s}$ (red dashed line) for comparison with the sound speed and pre-collision velocity ($v_{\rm c} = 1.0~c_{\rm s}$). We fit the $B-n$ relation with the power law $B\propto n^{j}$. The blue lines indicate the results of the least-squares fit. The power-law index of the $B-n$ relation can be used to verify the dynamic importance of a magnetic field \citep{2012ARA&A..50...29C}. For a dense gas in which the magnetic field is too weak to prevent isotropic gas collapse, the power-law index is expected to be 2/3 \citep{1966MNRAS.133..265M}, whereas for a stronger magnetic field, the power-law index decreases. Figure \ref{fig:n-B} shows that the magnetic fields in the three models were dominated by magnetic fields stronger than the initial strength $B_{0}$.

The left-hand panel of Figure \ref{fig:n-B} shows the results for the weak magnetic model {\tt b1-M1-By10}. The magnetic fields in the core were expected to have a minor effect on the gas motion because the Alfvén speed of most gases in the core was less than the sound speed (pre-collision speed). Gas mass generally exhibits a positive correlation between $B$ and $n$. The derived power-law index was 0.57.

The middle and right panels of Figure \ref{fig:n-B} show the results of the intermediate and strong magnetic models {\tt b1-M1-By30} and {\tt b1-M1-By50}, respectively. The estimated power-law indices were 0.37 and 0.12, respectively. For these models, the Alfvén speed of the gas in the core was comparable to the sound speed. Therefore, the magnetic fields in the core have a crucial effect on gas motion. The power-law index also demonstrates the major dynamic importance of the magnetic field. 

For {\tt b1-M1-By30} and {\tt b1-M1-By50}, unlike the weak magnetic model {\tt b1-M1-By10}, only one star formed. This can be explained by the higher magnetic Jeans mass of $M_{\rm J,mag} = M_{\rm J}(1+\beta^{-1})^{3/2}$ (e.g., \citealp{10.3389/fspas.2019.00007}), where $M_{\rm J}$ and $\beta$ are the purely thermal Jeans mass and the plasma $\beta$, respectively. As shown in Figure \ref{fig:n-B}, for these two models, Alfvén speeds of the gas were an order of magnitude higher than those for the model {\tt b1-M1-By10}. These higher Alfvén speeds increased the magnetic Jeans mass; therefore, the shocked layer did not fragment into the binary. 


\subsubsection{Effect of initial magnetic field orientation}

\begin{figure*}
    \begin{center}
    \includegraphics[clip,height=0.3\hsize]{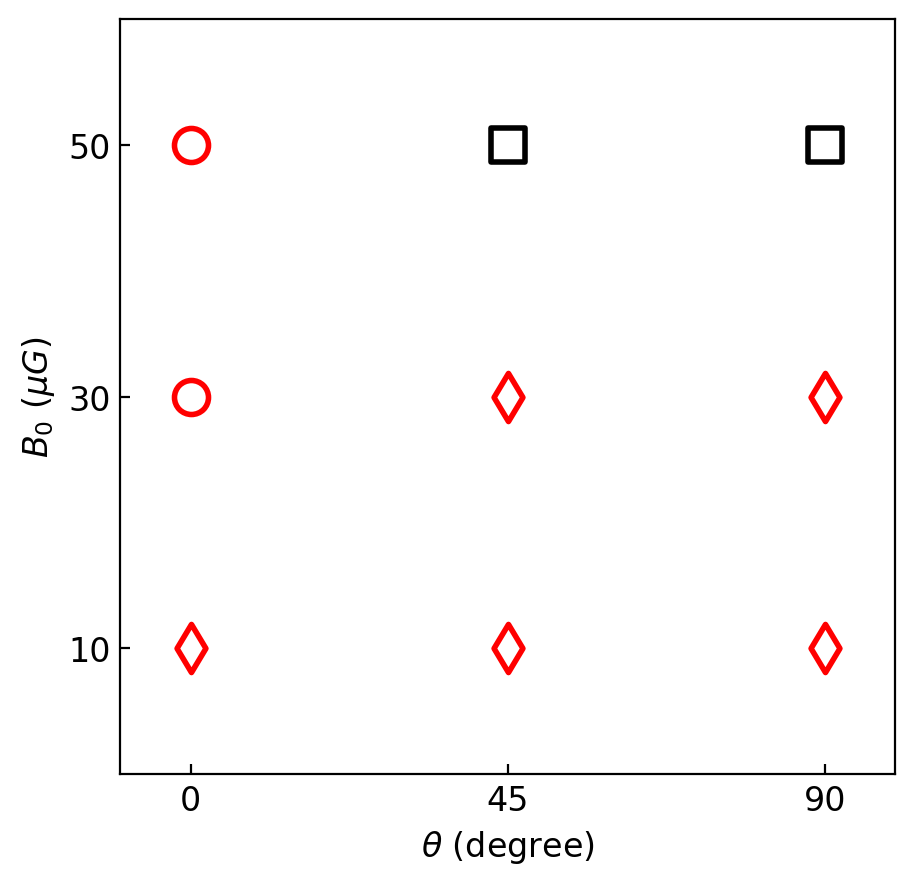}
     
    \vskip5pt  
    \caption{Simulation results in the $B_{0}$ vs. $\theta$ plane for models with $b = 1$ and $\mathcal{M} = 1$. Each red circle corresponds to a model that resulted in single star formation, each red diamond corresponds to a binary formation, and each black square corresponds to a merging without collapsing. 
\label{fig:SF_condition}}
    
\end{center}
\end{figure*}

For off-center collisions, owing to angular momentum, the direction of the magnetic field became more complicated, and the gas flow was significantly affected. 
Figure \ref{fig:SF_condition} shows the simulation results in the $B_{0}$ vs. $\theta$ plane for the models with $b = 1$ and $\mathcal{M} = 1$. For the $\theta = 0^{\circ}$ or $B_{0} < 50~\mu G$ cases, protostars formed, whereas when the magnetic field was strong ($B_{0} = 50~\mu G$) and $\theta \geq 45^{\circ}$, star formation was not induced and the cores merged. In other words, when the magnetic field component perpendicular to the collision axis is strong, star formation is not induced. The initial magnetic-field orientation plays an important role in the collision process and star formation.

For the $\theta = 0^{\circ}$ model, the magnetic field was parallel to the $xy$ plane. As shown in Figure \ref{fig:b1-M1-By30}, the magnetic field was bent parallel to the shock-compressed layer. The magnetic pressure does not work parallel to the magnetic field, and so the dense gas in the shock-compressed layer flowed into the center region along the direction of the magnetic field unimpeded by the magnetic pressure. A single star formed in the central region. The results are similar for model {\tt b1-M1-By50}.

In contrast, for the $\theta = 45^{\circ}$ model {\tt b1-M1-Bob30} and $\theta = 90^{\circ}$ model {\tt b1-M1-Bz30}, the initial magnetic field was nearly parallel to the $z$-axis. As shown in Figure. \ref{fig:b1-M1-Bz30},
the gas flow along the shock-compressed layer was more restricted than that in the $\theta = 0^{\circ}$ model. A large amount of gas did not accumulate in the center region and was binary. For the strong magnetic field model {\tt b1-M1-Bz50}, the gas did not accumulate (see Figure \ref{b1-M1-Bz50}), and the mass of the magnetic Jeans was higher. Therefore, star formation was not triggered. 

\begin{figure*}
    \begin{center}
    \includegraphics[clip,height=0.3\hsize]{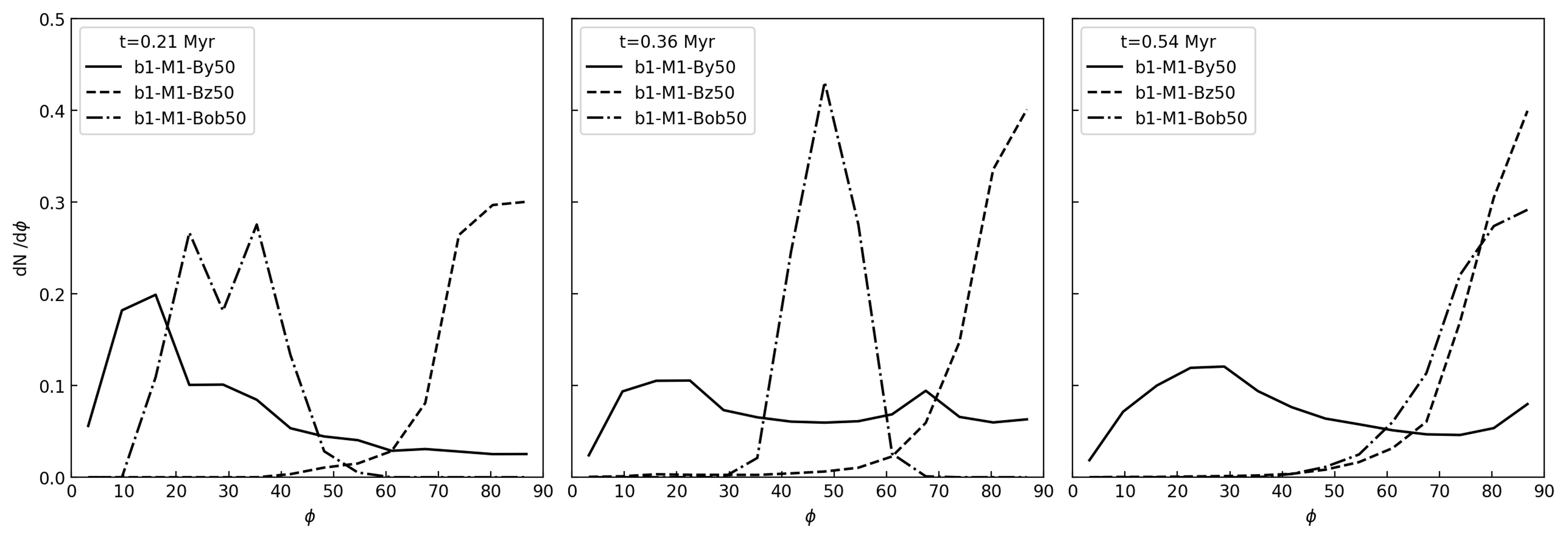}
     
    \vskip5pt  
    \caption{Histogram of the relative angle between the magnetic field $B$ and the velocity of dense gas ($ > 10^{5}$ cm$^{-3}$) pixel-by-pixel in models {\tt b1-M1-By50}, {\tt b1-M1-Bz50}, and {\tt b1-M1-Bob50}. Histograms at 0.21, 0.36, and 0.64 Myr are shown. Histograms with peaks at 90$^{\circ}$ correspond to $B$ predominantly perpendicular to the dense gas flow. 
\label{fig:v-B}}
    
\end{center}
\end{figure*}

To quantitatively investigate how the direction of the B-field correlated with gas motion, we quantified the degree of alignment of the magnetic field with respect to the gas velocity pixel-by-pixel. Figure 
\ref{fig:v-B} shows a histogram of the relative angle between the magnetic field $B$ and velocity of dense gas ($ > 10^{5}$ cm$^{-3}$) pixel-by-pixel in models {\tt b1-M1-By50}, {\tt b1-M1-Bz50}, and {\tt b1-M1-Bob50}. Histograms peaking at $\phi = 0^{\circ}$ indicate that B-fields were preferentially aligned parallel to the gas flow, whereas peaks at $\phi = 90^{\circ}$ indicate preferentially perpendicular alignment. We show the histograms at 0.21, 0.36, and 0.64 Myr. Throughout the simulation time, the gas flow in the $\theta = 0^{\circ}$ model {\tt b1-M1-By50} was preferentially aligned parallel to the magnetic field. Therefore, gas movement was relatively unimpeded by the magnetic pressure and gas tended to accumulate easily, leading to protostar formation. In contrast, the gas flow in the $\theta = 90^{\circ}$ model {\tt b1-M1-Bz50} was strongly perpendicular to the B-fields. Hence, magnetic pressure inhibited gas flow, and higher-density gas did not easily yield. In the $\theta = 45^{\circ}$ model {\tt b1-M1-Bob50}, the histogram initially peaked between $0^{\circ}$ and $90^{\circ}$, but eventually peaked at approximately $90^{\circ}$; therefore, the magnetic field pressure acts in directions parallel to the gas flow and hinders dense gas formation. Thus, a strong magnetic field perpendicular to the collision axis inhibits gas flow and suppresses star formation.

\subsection{Mixing of gas}
\label{sec:Mixing of gas}
In some models, star formation is triggered by dense core collisions. However, sometime after protostar formation, the two cores did not retain their original shapes. It is difficult to distinguish whether star formation is caused by core collision or contraction of a single core based only on the characteristics of the density distribution and magnetic field structure.
However, if pre-collision cores have different chemical compositions, we might observe the relics of the collision observationally, even after star formation. 

\begin{figure*}
    \begin{center}
    \includegraphics[clip,height=0.35\hsize]{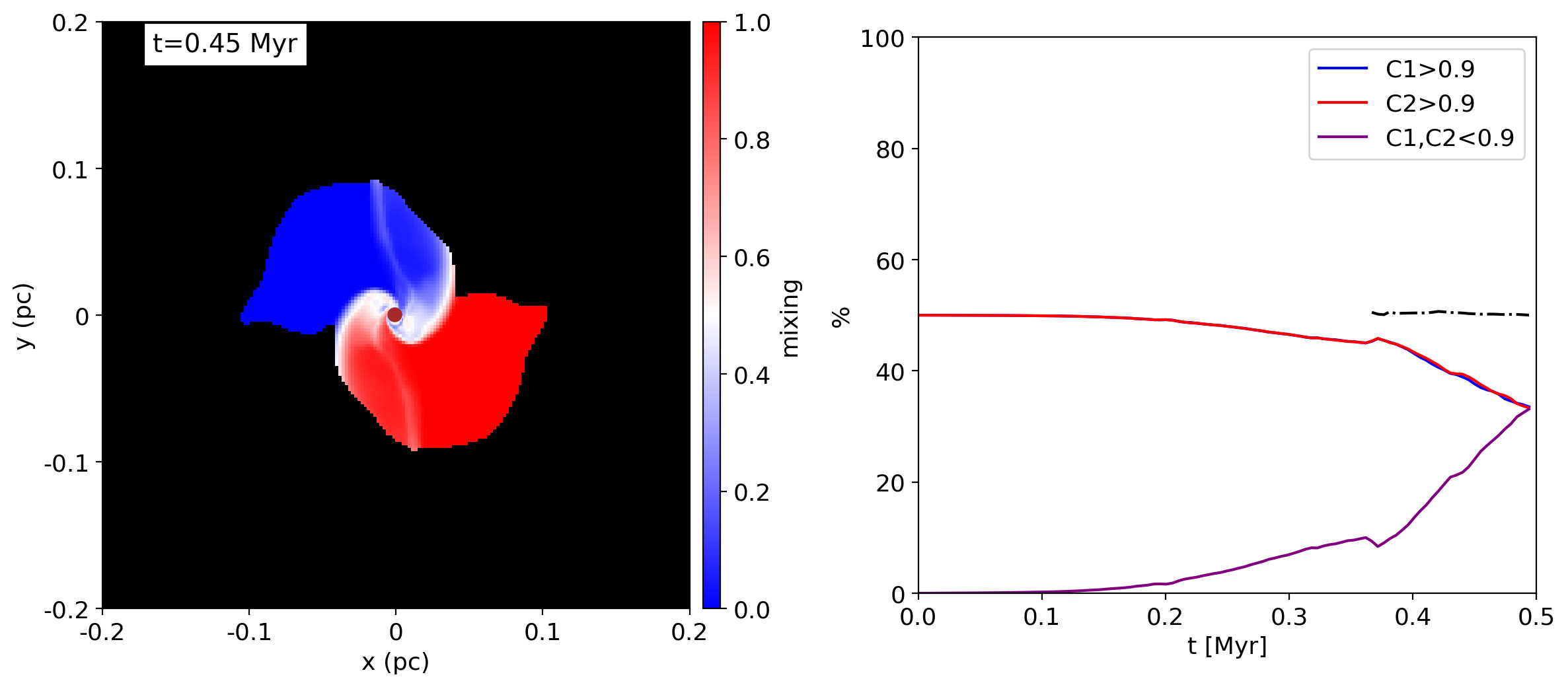}
     
    \vskip5pt  
    \caption{\textit{Left}: Ration of the integration in the line of sight represented by $\int_{C_{1}+C_{2} > 0.5} \rho C_{1}dx/\int_{C_{1}+C_{2} > 0.5} \rho (C_{1}+C_{2})dx$ for model {\tt b1-M1-By30} at $t = 0.45$ Myr. For cells with ${C_{1}+C_{2} > 0.5}$, the integration in the line of sight of $\rho C_{1}$ was divided by that of $\rho (C_{1}+C_{2})$. This is an indicator of how much gas is being mixed. The purple point represents the sink particle. The black region indicates where $\int_{C_{1}+C_{2} > 0.5} \rho (C_{1}+C_{2})dx = 0$ \textit{Right}: Mass fraction of the gas. The blue line indicates the mass fraction of the gas where $C_{1} > 0.9$ in the whole gas. The red line indicates the mass fraction of gas where $C_{2} > 0.9$. The purple line indicates that of the gas where both $C_{1} < 0.9$ and $C_{2} < 0.9$. The black dashed-dot lines indicate the mass fraction in the sink particle of the gas that is initially labeled $C_{1} = 1.0$. 
\label{fig:mixng-b1-M1-By30.png}}
    
\end{center}
\end{figure*}

The left panel in Figure \ref{fig:mixng-b1-M1-By30.png} shows the ratio of integration in the line of sight, represented by: 
\begin{equation}
 \label{eq:mixing integration}
 \rm{integrated~ mixing~ rate} = \frac{\int_{C_{1}+C_{2} > 0.5}\rho C_{1}dx}{\int_{C_{1}+C_{2} > 0.5} \rho (C_{1}+C_{2})dx,}
\end{equation}
for model {\tt b1-M1-By30} at $t = 0.45$ Myr (see section \ref{sec:color}). 
This integration is a barometer of how much gas from each core is mixed; values close to 0 indicate a large amount of gas originating from one core, while a value close to 1 indicates a large amount originating from the other core; values around 0.5 means that the two are equally mixed. At the collision interface, the two gases mixed well, whereas in spiral arm structures, the contribution of one gas was larger. The arms and discs around the protostar had almost the same gas composition as that in the initial state. The right panel of Figure \ref{fig:mixng-b1-M1-By30.png} shows the mass fraction of the gas. As the collision proceeded, the fraction of mixed gas increased; however, for approximately 0.1 Myr after particle formation, the gas mass with $C_{1} > 0.9$ or $C_{2} > 0.9$ was large. Thus, if we observe a notable abundance anomaly in the molecular envelope surrounding a star, it could suggest that the system was formed by the collision process of the two cores with different abundances.

\begin{figure*}
    \begin{center}
    \includegraphics[clip,height=0.5\hsize]{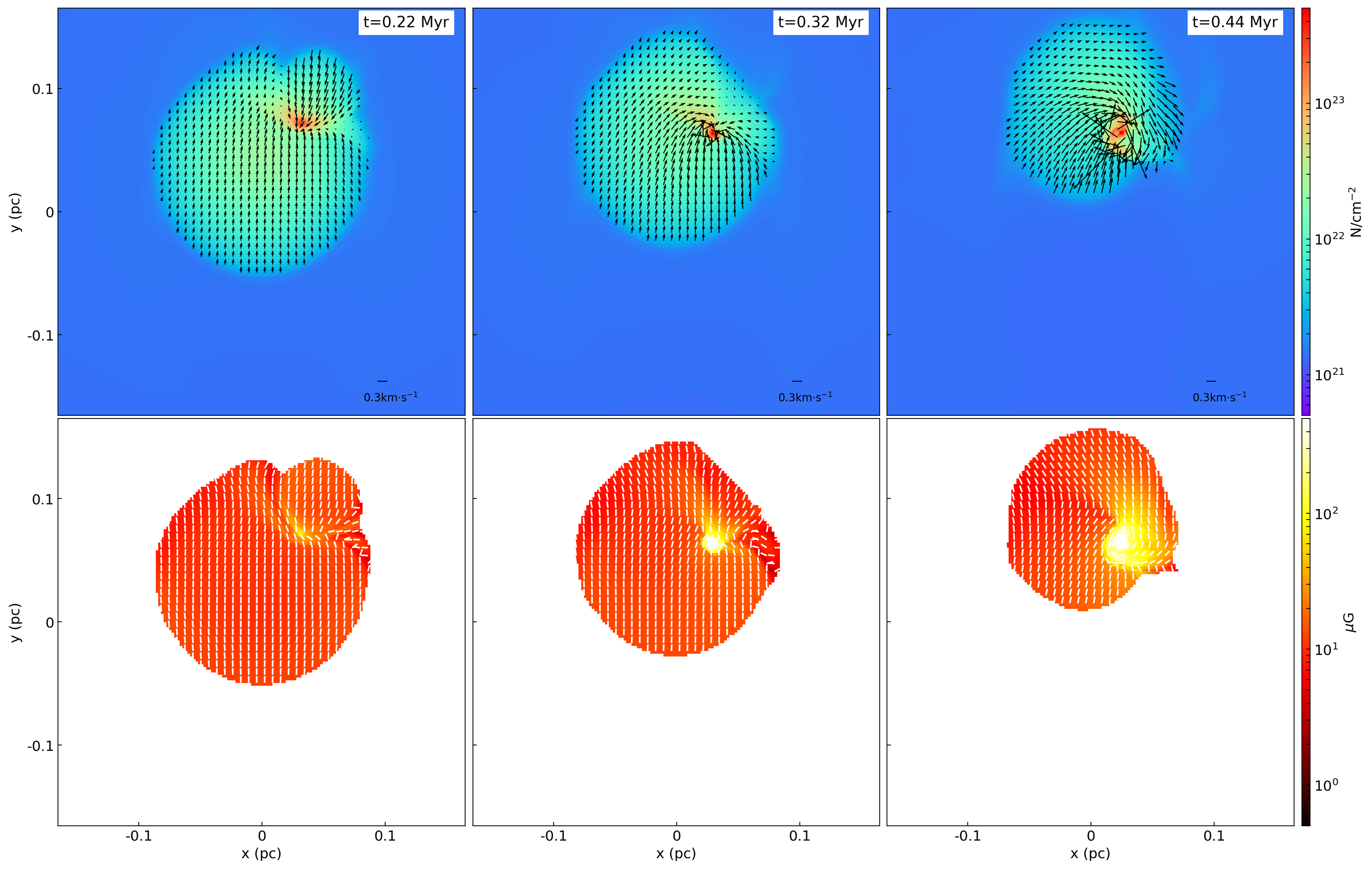}
     
    \vskip5pt  
    \caption{As in Figure \ref{fig:b0-M1-By10} for the collision of unequal-mass cores. Snapshots at 0.22, 0.32, and 0.44 Myr are shown. 
\label{fig:dissimilar_mass_slice}}
    
\end{center}
\end{figure*}

\begin{figure*}
    \begin{center}
    \includegraphics[clip,height=0.5\hsize]{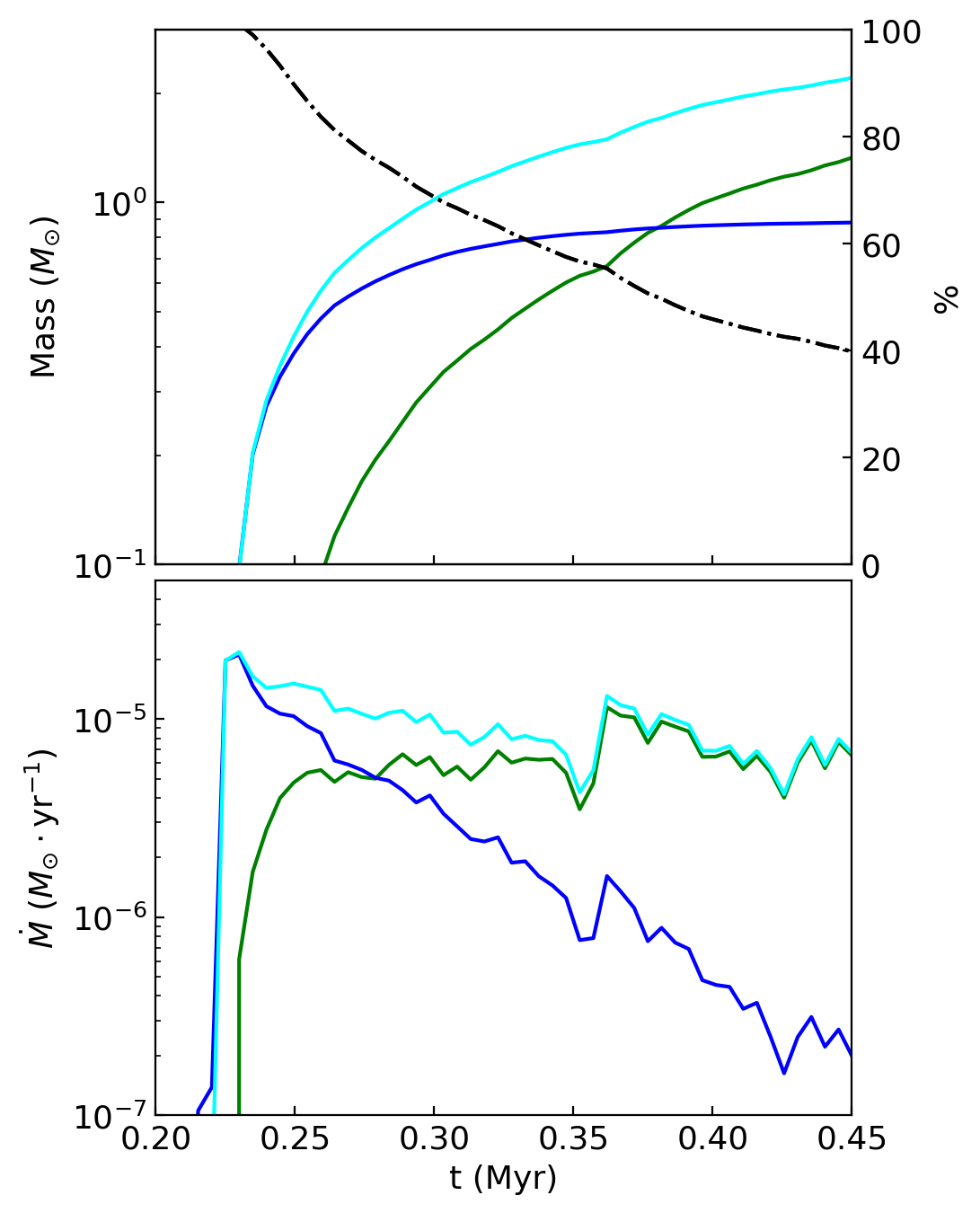}
     
    \vskip5pt  
    \caption{Evolution of mass (top) and mass accretion rate (bottom) of the sink particle for the collision of unequal-mass cores. The blue lines correspond to the contribution from gas initially labeled $C_{1} = 1.0$. The green lines correspond to that from gas labeled $C_{2} = 1.0$. The cyan lines indicate the total amount. In the top row, the black dashed-dot lines indicate the mass fraction of gas initially labeled $C_{1} = 1.0$ for the sink particle. 
\label{fig:dissimilar_mass_rate}}
    
\end{center}
\end{figure*}

To further investigate the gas mixing process, we briefly examined additional models of the collision of unequal mass cores. In these models, we initially prepared two stable BE spheres at a mass ratio of 4:1. The first was the same condition as the previous models, with a radius of $r_{\rm c1} = 0.1 ~\rm pc$, central density of $\rho_{\rm c1} = 10^{5} ~\rm cm^{-3}$, temperature of $T_{\rm c1} = 20 ~\rm K$, and mass of $M_{\rm c1} = 3.7 ~M_{\odot}$. The other was smaller and lighter with a radius of $r_{\rm c2} = 0.05 ~\rm pc$, central density of $\rho_{\rm c2} = 2.0\times10^{5} ~\rm cm^{-3}$, temperature of $T_{\rm c2} = 10 ~\rm K$, and mass of $M_{\rm c2} = 1.9 ~M_{\odot}$. The gas components initially contained in the smaller and larger cores are labeled with color variables $(C_{1},C_{2}) = (1,0)$ and $(0,1)$, respectively. By contrast, the ambient gas is labeled $(C_{1},C_{2}) = (0,0)$.
For both cores, the pressure balance at the core boundary was satisfied. The cores had a pre-collision velocity of $v_{\rm c} = 0.19 ~\rm km~s^{-1}$, which corresponds to the sound speed in the smaller core with the impact parameter $b = r_{\rm c2}$. Initially, we imposed a uniform magnetic field $B_{\rm y} = 10 ~\mu G$, which was parallel to the collision direction. Figure \ref{fig:dissimilar_mass_slice} shows snapshots of the collision. Figure \ref{fig:dissimilar_mass_rate} shows the evolution of the mass and mass accretion rates of the sink particle. The contribution from each core is indicated based on an analysis using color variables.
Initially, the smaller core plunged into the larger core, and a compressed layer formed in front of the core. At t = 0.22 Myr, the sink particle formed and the gas components of the smaller core accreted onto the particle with $\dot{M}\sim ~10^{-5}M_{\odot}~\rm{yr}^{-1}$. 
Larger core components proceeded in the $y$-positive direction. At approximately t = 0.3 Myr, the accretion rate from the smaller core components decreased to $\dot{M}\sim 10^{-6}~M_{\odot}~\rm{yr}^{-1}$, followed by secondary accretion from the larger core components. Larger core components accreted and rotated around the particle. As shown in the rightmost panel, a one-arm spiral pattern resembling a whirlpool formed. The polarization vectors followed this spiral stream. In this way, for the collision of unequal mass, one core collapsed first, and the gas from the other core accreted next. This may have caused isotopic anomalies similar to those observed in our solar system \citep{1973Sci...182..485C}. 


\section{Conclusions}
\label{sec:conclusion}
We conducted a series of 3D MHD simulations to study the evolution of dense-core collisions in a magnetic field. The core was modeled as a stable isothermal Bonnor–Ebert sphere. Adaptive mesh refinement and sink-particle techniques were used to follow the evolution of colliding cores. We explored the parameter space of dense core collisions, including the offset parameter $b$, Mach number of the initial core $\mathcal{M}$, magnetic field strength $B_{0}$, and angle $\theta$ between the initial magnetic field and collision axis. Below, we summarize the primary findings. 
\begin{quote}
 \begin{itemize}
  \item For the case of a head-on ($b = 0$) collision, one protostar formed. As the initial magnetic field strength was higher (with $\mathcal{M}$ and $\theta$ held constant), the rate of increase in density and the accretion rate onto the protostar decreased. 
  As magnetic pressure acts in directions perpendicular to the field lines, for models with $\theta = 90^{\circ}$, the accretion rate was more dependent on the initial magnetic field strength, as compared with $\theta = 0^{\circ}$ models. The initial state of the magnetic field had a strong effect on the mass growth of the protostar. 
  
  \item In the case of an off-center ($b = 1$) collision, the gas had an angular momentum, and therefore the structures were complicated. The motion of gas and number of protostars strongly depended on the initial magnetic field strength and orientation. The dynamic importance of the magnetic field determined the magnetic Jeans mass, and therefore the initial magnetic field strength affected the fragmentation of dense gas and the number of protostars. Besides, the evolution of the collision process depended on $\theta$, because the magnetic field orientation influences gas motion. Even in the case of an off-center collision, the initial state of the magnetic field affected the results. 
  
  \item In our simulations, we use the color variable to track the mixing of the gas. We found that the gas components of two cores mixed well at the collision interface, while in the arm structures the mixing rate was very low 0.1 Myr after star formation. If pre-collision cores have different chemical compositions, we may find observational relics of the collision by investigating the abundance of gas.
\end{itemize}
\end{quote}

\clearpage
\bibliography{sample631}{}
\bibliographystyle{aasjournal}



\end{document}